\documentclass[showpacs,preprintnumbers,prd,nofootinbib,floats,amssymb,floatfix]{revtex4}
\usepackage{amsmath}
\usepackage{amsfonts}
\usepackage{graphicx}
\usepackage{hyperref}

\usepackage{amsmath}
\usepackage{amstext}
 
\begin{document}
\title{Hamiltonian dynamics and  Faddeev-Jackiw quantization of  3D  gravity with a Barbero-Immirzi like parameter }
\author{Alberto Escalante}  \email{aescalan@sirio.ifuap.buap.mx}
 \affiliation{  Instituto de F{\'i}sica, Universidad Aut\'onoma de Puebla,  \\
 Apartado Postal J-48 72570, Puebla Pue., M\'exico. }
\author{J. Manuel-Cabrera } \email{}
 \affiliation{  Instituto de F{\'i}sica, Universidad Aut\'onoma de Puebla,  \\
 Apartado Postal J-48 72570, Puebla Pue., M\'exico, }

\begin{abstract}
A detailed Dirac's and  Faddeev-Jackiw quantization of   Bonzom-Livine model describing  gravity in three dimensions is performed. The full structure of the constraints,  the gauge transformations and the generalized Faddeev-Jackiw brackets are found. In addition,  we show that the Faddeev-Jackiw  and  Dirac's brackets coincide to each  other. Finally we discuss some remarks and prospects.
\end{abstract}
 \date{\today}
\pacs{98.80.-k,98.80.Cq}
\preprint{}
\maketitle
\section{INTRODUCTION}
It is well-know that 3-dimensional gravity is  an interesting toy  model. In fact,  it is considered  as good test theory   for trying to   understand  the difficulties   that emerge in the quantization of four dimensional gravity. It is worth to mention, that since  the works developed by Achucarro, Towsend, Witten  and  other authors \cite{1, 2},  there is  a huge effort for understanding at classical and quantum level   the connection between gravity and gauge connections  theories just like Chern-Simons theory \cite{3},  then, it  is expected that  the learned   in the three dimensional case  could  be useful for  constructing  better tools and  apply it  for  the quantization   of   four-dimensional gravitational theory. In this respect,  it is common to obtain in three dimensions a relation  between   Palatini   and Chern-Simons theory, in fact, it has been showed that these theories are equivalent  up to a total derivative \cite{2, 3, 4}. However,  the relation reported between these theories  is not the only one,    there is a new   action classically equivalent  to Palatini's theory,   it  is the   so-called exotic action with a Barbero-Immirzi like parameter  \cite{5} (we call it from now on Bonzom-Livine action [BL] ). In fact,  [BL]  model describes a set of actions sharing   the same  equations of motion with Palatini's theory, however, the  symplectic structure is different to each other. The symplectic structure  in [BL] model depends of a Barbero-Immirizi like parameter, from which, one expects  that  the  quantum theories will be different \cite{4}. In this respect, something similar happens in the four dimensional case with the Holst action \cite{6}. The Holst action provides  a  set of  actions classically equivalent to Einstein's   theory,  it depends of a parameter called Barbero-Immirizi (we  call it $\gamma$  parameter) and the contribution of this  parameter can be appreciated   at classical level  in the symplectic structure of the theory    and  the  coupling of fermionic matter with gravity, in fact, it  determines the coupling constant of a four-fermion interaction \cite{7}.  From the quantum point of view, the $\gamma$ parameter gives  a contribution in the quantum  spectra of the   area and volume  operators  in the Loop Quantum Gravity context \cite{8}. Furthermore, the term added by Holst  to Palatini's action facilitates so much the canonical description of General Relativity,  and depending of the values of   $\gamma$   we can reproduce  the different scenarios found in canonical gravity, for instance, it is possible to obtain the ADM, Ashtekar and Barbero formulations in  straightforward way \cite{6}.   Nevertheless, in spite of the Holst  action provide a general action for gravity, the $\gamma$ parameter  is still controversial \cite{8}. In this manner,  the [BL] action becomes to be the three dimensional  equivalent  model to Holst's  action, in fact, the equivalence is not given only with   the presence of an $\gamma$ parameter, but  also  at classical level  if we  perform  a partial gauge fixing  in the canonical description of [BL], then  it is possible to obtain  a full Ashtekar's  connection dynamics   in three dimensions \cite{4}. In this respect, the analysis of the symmetries of the  [BL] action   has been performed in \cite{4, 5},  in these works  the canonical  analysis by using the Dirac method was performed. However, in these works the analysis was developed  on a smaller phase space and the complete structure of the constraints on the full phase space was not reported. It is important to remark that if some of the Dirac steps is omitted, then it is possible to obtain incomplete results \cite{9, 10}. In this manner, an  analysis developed on the full phase space and following all the Dirac steps is mandatory. However, in some cases,  to develop   the Dirac method for gauge theories  is   large and tedious task, hence, because of  these complications,   it is necessary to use alternative formulations that could give us a complete canonical description of the theory, in this sense,    there is a different  approach  for studying  gauge theories, it is  called  the Faddeev-Jackiw  [FJ] formalism \cite{11}. The [FJ] method   is a symplectic approach, namely, all the relevant information of the theory can be obtained through an invertible symplectic tensor, which is  constructed by means  the symplectic variables that are identified as the degrees of freedom.  Because of the theory under interest is singular  there will be constraints,  and [FJ] has the   advantage that  all the  constraints of the theory are at the same footing, namely,   it is not necessary  to perform  the  classification of the constraints in primary, secondary, first class or second class  as in Dirac's method is done \cite{12}. When  the symplectic tensor is obtained, then  its components are identified with the [FJ]  generalized  brackets, Dirac's brackets and [FJ] brackets coincide to each other. \\
Because of the explained above, in this paper we develop a pure Dirac's method and a  full [FJ] analysis of the [BL] model. In fact, in order to compare both approaches,  it is necessary to work   in Dirac's method with the complete configuration space. Hence,  for constructing the Dirac brackets  and compare it with the generalized [FJ] ones,  we need to know the complete structure of the constraints over the full phase space \cite{13}. Furthermore, we shall prove that the [FJ] approach is more economic than Dirac's one. It is important to comment that our results has not been reported in the literature and as special case we reproduce those reported in \cite{4, 5}. In addition, we would also  remark  that for [BL] theory we shall construct  the Dirac brackets by eliminating the second class constraints and remaining the first class ones. Furthermore, at the end of the paper,  we have added an appendix where the analysis of an Abelian  [BL]   theory is performed, in that appendix, we construct the Dirac brackets by fixing the gauge and also we reproduce all those results by means of [FJ] formalism. \\
The paper is organized as follows;  in Section II a detailed canonical analysis of [BL] is performed. We report the complete structure of the constraints defined on the full phase space, then we eliminate the  second class constraints by constructing  the Dirac brackets. In Section III,  we study the relation between [BL]  and  Chern-Simons theory. We reproduce the results of the previous section by performing a pure Dirac's analysis of a generalized Chern-Simons theory.  In the Section IV, a detailed [FJ] of [BL] action is developed.  In order to reproduce all the Dirac results, we work with the configuration space field as symplectic variables, we  identify all the constraints of the theory and we show that the [FJ] generalized and Dirac's brackets coincide each to other. In Section V we add some remarks and conclusions.   \\

\section{ Hamiltonian  dynamics for three dimensional BL  gravity }
In this section, we will study the Hamiltonian dynamics of the action proposed by  [BL] \cite{5}. We will perform our analysis by using a pure Dirac's method, namely, we will find all the constraints defined on the full phase space. As was comment above, there is  an  analysis of the [BL] action developed on a smaller phase space  reported in \cite{4,5}, however, in those works the structure of the constraints is not  complete, thus, in order to compare  the [FJ] method with the Dirac one  it is mandatory to perform the Dirac  analysis on the full phase space by  following  all the Dirac steps.\\
It is well-known  that three dimensional gravity with a cosmological constant can be written  as a Chern-Simons theory \cite{1, 2, 3, 4, 5}. In fact, if the principal gauge bundle $G$ over $M$ is given by $G=SU(2)$ for 3d Euclidean gravity,  then we  can enlarge the group $G$ to $\tilde G$, where $\tilde G$ could be $SO(4)$, $ISO(3)$ or $SO(3,1)$ depending on the sign of the cosmological constant $\Lambda$  positive, zero or negative respectively. Hence, the algebra of  the generators of $\tilde G$ will satisfy the following commutation relations \cite{4, 5}
\begin{equation}
[ J_i, J_j ]= \epsilon_{ij}{^{k}}J_k \quad \quad \quad  [ J_i, K_j ]= \epsilon_{ij}{^{k}}K_k \quad \quad \quad  \left[ K_i, K_j \right]=s \epsilon_{ij}{^{k}}J_k,
\end{equation}
where $s=-1, 0, 1$, corresponding to the sign of the  cosmological constant and $i, j, k=1, 2, 3$. In order to construct a Chern-Simons theory being equivalent to standard Einstein's action of gravity, we choose the following non-degenerate invariant bilinear form
\begin{equation}
\left<  J_i, K_j \right> =\delta_{ij}, \quad \quad  \left<  J_i, J_j \right> =\left<  K_i, K_j \right>=0,
\end{equation}
in this manner, 3d Palatini's  action  with  cosmological constant can be written as
\begin{equation}
S'_{Palatini}= \frac{1}{\sqrt{|\Lambda|}} \int_M \epsilon^{\mu \nu \rho} \left( \left< A_\mu, \partial_\nu A_\rho \right> + \frac{1}{3} \left< A_\mu, [A_ \nu, A_\rho] \right> \right).
\end{equation}
On the other hand, if $\Lambda \neq 0$, then  there is another invariant non-degenerate bilinear form given by
\begin{equation}
(J_i, J_j)=\delta_{ij} \quad \quad \quad (K_i, K_j)=s\delta_{ij} \quad \quad \quad (J_i, K_j)=0,
\end{equation}
in this case, we can obtain from the following Chern-Simons action
\begin{equation}
\tilde S_{Exotic} = \frac{1}{\sqrt{|\Lambda|}} \int_M \epsilon^{\mu \nu \rho} \left( \left(A_\mu, \partial_\nu A_\rho \right) + \frac{1}{3} \left( A_\mu, [A_ \nu, A_\rho] \right) \right),\end{equation}
the so-called exotic action for gravity
\begin{equation}
\widetilde{S}[A,e]=\frac{1}{\sqrt{\mid\Lambda\mid}}\left[ \int_M A^{i} \wedge d A_{i} + \frac{1}{3}\epsilon_{ijk}A^{i}\wedge A^{j} \wedge A^{k}\right]+ s\sqrt{\mid\Lambda\mid}\int_M  e^{i}\wedge d_{A}e_{i},
\end{equation}
where the 1-form  $A^{i}=A{_{\mu}}^{i}dx^{\mu}$,  $(d_{A}v)^{i}= dv^{i}+ [A,v]^{i}=dv^{i}+\epsilon{^{i}}_{jk}A^{j}\wedge v^{k} $ and $ F^{i}=dA^{i} +\frac{1}{2} \epsilon^{i}{_{jk}}A{^{j}} \wedge A{^{k}}$ is the strength  two-form. In this manner, the [BL] model consists of  considering  the combination of Palatini's  and exotic action through a parameter, namely $\gamma$, being a kind of  Barbero-Immirizi  parameter
\begin{equation}
S{_{\gamma}}[A,e]=S_{Palatini}'[A,e]+\frac{1}{\gamma}\widetilde{S}_{Exotic}[A,e].
\label{eq1a}
\end{equation}
In fact, from the  action (\ref{eq1a}) we obtain a family of theories classically equivalent to 3d gravity  in the sense that Palatini's theory with cosmological constant and [BL] actions share the same equations of motion, which  can be seen from the variation of the action (\ref{eq1a})
\begin{equation}
\frac{\delta S{_{\gamma}}[A,e]}{ \delta e_{\mu i}}: \qquad  \epsilon^{\mu\nu\rho}\left[ F{^{i}}_{\nu\rho}[A]+s\frac{\mid\Lambda\mid}{2} \epsilon{^{i}}_{jk}e{_{\nu}}^{j} e{_{\rho}}^{k}\right]+s\frac{\sqrt{\mid\Lambda\mid}}{\gamma}\epsilon^{\mu\nu\rho} D_\nu e{^{i}}_{\rho}=0,
\label{eq4}
\end{equation}
\begin{equation}
\frac{\delta S{_{\gamma}}[A,e]}{ \delta A_{\mu i}}: \qquad \epsilon^{\mu\nu\rho}D_\nu e{^{i}}_{\rho}+\frac{1}{\gamma\sqrt{\mid\Lambda\mid}}\epsilon^{\mu\nu\rho}\left[ F{^{i}}_{\nu\rho}[A]+s\frac{\mid\Lambda\mid}{2} \epsilon{^{i}}_{jk}e{_{\nu}}^{j} e{_{\rho}}^{k}\right]=0,
\label{eq5}
\end{equation}
the equations (\ref{eq4}) and (\ref{eq5}) are equivalent to Einstein's equations.  Hence, in order to develop the Hamiltonian analysis, we perform the 2+1 decomposition of the action  (\ref{eq1a}) obtaining
\begin{eqnarray}
S{_{\gamma}}[e,A]&=& \int d^{3}x \bigg[ 2\epsilon^{0ab}\delta_{ij}(e{_{0}}^{i}+\frac{1}{\gamma\sqrt{\mid\Lambda\mid}}A{_{0}}^{i})(F^{j}{_{ab}}+s\frac{\mid\Lambda\mid}{2}\epsilon{^{j}}_{kl}e^{k}_{a}e^{l}_{b})+2\epsilon^{0ab}\delta_{ij}D_{a}e{_{b}}^{i}(A^{j}_{0}+s\frac{\sqrt{\Lambda}}{\gamma}e^{j}_{0})\nonumber\\ &\,&+ 2\epsilon^{0ab}\delta_{ij}(e^{i}_{b}\partial_{0}A^{j}_{a}+\frac{1}{2\gamma\sqrt{\mid\Lambda\mid}}A^{i}_{b}\partial_{0}A^{j}_{a}+s\frac{\sqrt{\mid\Lambda\mid}}{2\gamma}e^{i}_{b}\partial_{0}e^{j}_{a})\bigg] ,
\label{eq6}
\end{eqnarray}
where $a, b, c= 1, 2$. The definition of the momenta $(\pi^{\alpha}{_{i}},\Pi^{\alpha}{_{i}})$ canonically conjugate to $( e{^i}_{\alpha}, A_{\alpha} {^{i}})$ is given by
\begin{equation}
\Pi^{\alpha}{_{i}}= \frac{\delta {\mathcal{L}} }{ \delta \dot{A}_{\alpha} {^{i}} },  \qquad \pi^{\alpha}{_{i}}= \frac{\delta {\mathcal{L}} }{ \delta \dot{e}{^i}_\alpha } .
\label{eq8}
\end{equation}
The matrix elements of the Hessian
\begin{equation}
 \frac{\partial^2{\mathcal{L}} }{\partial (\partial_\mu  e^{i}_\alpha) \partial (\partial_\mu e^{i}_\beta )} , \quad \frac{\partial^2{\mathcal{L}} }{\partial( \partial_\mu  e^{i}_\alpha) \partial (\partial_\mu A_{\beta} {^{i}} )}, \quad \frac{\partial^2{\mathcal{L}} }{\partial (\partial_\mu A_{\alpha} {^{i}} ) \partial(\partial_\mu A_{\beta} {^{i}} ) },
\label{eq9}
\end{equation}
are identically zero, thus, we expect $18$ primary constraints. From the definition of the momenta $(\ref{eq8})$ we identify the following  $18$ primary constraints
\begin{eqnarray}
\phi{_{i}}^{0}&:=& \pi{_{i}}^{0} \approx 0 ,\nonumber \\
\phi{_{i}}^{a}&:=& \pi{_{i}}^{a}- s\frac{\sqrt{\mid\Lambda\mid}}{\gamma}\epsilon^{0ab}\delta_{ij} e{_{b}}^{j} \approx 0 ,\nonumber \\
\Phi{_{i}}^{0}&:=& \Pi{_{i}}^{0} \approx 0 ,\nonumber \\
\Phi{_{i}}^{a}&:=& \Pi{_{i}}^{a}- 2\epsilon^{0ab}\delta_{ij}(e{_{b}}^{j}+\frac{1}{2\gamma\sqrt{\mid\Lambda\mid}}A{_{b}}^{j})  \approx 0.
\label{eq10}
\end{eqnarray}
The canonical Hamiltonian takes the form
\begin{equation}
H_c= \int dx^2 \left[-2\epsilon^{0ab}\delta_{ij} D_{a}e{^{i}}_{b}(A^{i}_{0}+s\frac{\sqrt{\Lambda}}{\gamma}e^{i}_{0})-2\epsilon^{0ab}\delta_{ij}(e{_{0}}^{i}+\frac{1}{\gamma\sqrt{\mid\Lambda\mid}}A{_{0}}^{i})(F^{j}{_{ab}}+s\frac{\mid\Lambda\mid}{2}\epsilon^{j}{_{kl}}e^{k}_{a}e{^{l}}_{b}) \right],
\label{eq11}
\end{equation}
and the primary Hamiltonian is given by
\begin{equation}
H_P= H_c + \int dx^2 \left[  \lambda^{i}{_{\alpha}} \phi{_{i}}^{\alpha} + \xi^{i}{_{\alpha}} \Phi{_{i}}^{\alpha}  \right],
\label{eq12}
\end{equation}
where $ \lambda^{i}{_{\alpha}}, \xi^{i}{_{\alpha}}$ are Lagrange multipliers enforcing the constraints $(\phi{_{i}}^{\alpha}, \Phi{_{i}}^{\alpha})$. The fundamental Poisson brackets of the theory  are given
\begin{eqnarray}
\{ e_{\alpha} {^{i}}(x),\pi^{\beta}{_{j}}(y) \}  &=& \delta {^ \beta} _\alpha \delta {^ i} _j \delta^2(x-y), \nonumber \\
\{ A_{\alpha} {^{i}}(x),\Pi^{\beta}{_{j}}(y) \} &=&  \delta {^ \beta} _\alpha \delta {^ i} _j \delta^2(x-y),
\label{eq13}
\end{eqnarray}
where we can observe that in these fundamental brackets there is not any contribution of the $\gamma$ parameter; in the Dirac brackets, however, there will be a non trivial contribution. In order to observe the presence of  more constraints, we calculate the following 18$\times$18 matrix whose entries are the Poisson brackets among the constraints (\ref{eq10})
\begin{eqnarray}
\{\phi{_{i}}^{a} (x),\phi{_{j}}^{b} (y) \} &=& -2s\frac{\sqrt{\mid\Lambda\mid}}{\gamma}\epsilon^{0ab} \delta_{ij}\delta^2(x-y), \nonumber \\
\{\phi{_{i}}^{a} (x),\Phi{_{j}}^{b} (y) \} &=& -2\epsilon^{0ab}\delta_{ij}\delta^2(x-y),\nonumber \\
\{\Phi{_{i}}^{a} (x),\Phi{_{j}}^{b} (y) \} &=& -2\frac{1}{\gamma\sqrt{\mid\Lambda\mid}}\epsilon^{0ab}\delta_{ij}\delta^2(x-y),
\label{eq14}
\end{eqnarray}
we appreciate that this matrix has rank=12 and 6 null-vectors. By using the 6 null-vectors  and consistency conditions we obtain the following   6 secondary constraints
\begin{eqnarray}
\gamma{_{i}}^{0}&=&   \pi{_{i}}^{0} \approx 0,  \nonumber \\
\tilde \gamma{_{i}}^{0}&=&   \Pi{_{i}}^{0}  \approx 0,  \nonumber \\
\dot{\phi}{_{i}}^{0}&=& \{\phi{_{i}}^{0} (x), {H}_{P} \} \approx 0 \quad \Rightarrow \quad \psi_{i}:= 2\epsilon^{0ab}s\frac{\sqrt{\mid\Lambda\mid}}{\gamma} D_{a}e_{ib}+2\epsilon^{0ab}(F_{iab}+\frac{s\mid\Lambda\mid}{2}\epsilon{_{ijk}}e{^{j}}_{a}e{^{k}}_{b})\approx 0, \nonumber \\
\dot{\Phi}{_{i}}^{0}&=& \{\Phi{_{i}}^{0} (x), {H}_{P} \} \approx 0 \quad \Rightarrow \quad \Psi_{i}:=2\epsilon^{0ab} D_{a}e_{ib}+2\epsilon^{0ab}\frac{1}{\gamma\sqrt{\mid\Lambda\mid}}(F_{iab}+\frac{s\mid\Lambda\mid}{2}\epsilon{_{ijk}}e^{j}{_{a}}e^{k}{_{b}})  \approx 0, \nonumber \\
\label{eq15}
\end{eqnarray}
and the rank allows us to fix the following  values for the Lagrangian multipliers
\begin{eqnarray}
\dot{\phi}{_{i}}^{a}&=&\{\phi{_{i}}^{a},H_{P}\}\approx0 \Rightarrow 2\epsilon^{0ab}\frac{s-\gamma^{2}}{\gamma}\sqrt{\mid\Lambda\mid}(-\lambda_{bi}+D_{b}e_{0i}+\epsilon_{lim}e^{m}_{b}A^{l}_{0})\approx0, \nonumber \\
\dot{\Phi}{_{i}}^{a}&=&\{\Phi{_{i}}^{a},H_{P}\}\approx0 \Rightarrow 2\epsilon^{0ab}\frac{s-\gamma^{2}}{\gamma^{2}}(-\xi_{bi}+D_{b}A_{0i}+s\mid\Lambda\mid\epsilon_{lim}e^{m}_{b}e^{l}_{0})\approx0.
\label{eq16}
\end{eqnarray}
Consistency requires conservation in time   of the secondary constraints, however, for  this theory there are not third constraints. At this point,  we need to identify from  the primary and secondary constraints which ones correspond to  first and second class. For this aim,  we need to calculate the rank and the null-vectors of the  24$\times$ 24 matrix whose entries will be the Poisson brackets between primary and secondary constraints, this is
\begin{eqnarray}
\{\phi{_{i}}^{a} (x),\phi{_{j}}^{b} (y) \} &=& -2s\frac{\sqrt{\mid\Lambda\mid}}{\gamma}\epsilon^{0ab}\delta_{ij}\delta^{2}(x-y), \nonumber \\
\{\phi{_{i}}^{a} (x),\Phi_{j} (y)\} &=& -2\epsilon^{0ab}\delta_{ij}\delta^{2}(x-y),\nonumber \\
\{\phi{_{i}}^{a} (x),\psi{_{j}}(y) \} &=&-2\epsilon^{0ab}\left[\frac{s\sqrt{\mid\Lambda\mid}}{\gamma}\delta_{ij}\partial_{b}-\epsilon_{ijk}(\frac{s\sqrt{\mid\Lambda\mid}}{\gamma}A{^{k}}_{b}+s\mid\Lambda\mid e{^{k}}_{b})\right] \delta^2(x-y) , \nonumber \\
\{\phi{_{i}}^{a} (x), \Psi_{j}(y) \} &=& -2\epsilon^{0ab}\left[\delta_{ij}\partial_{b}-\epsilon_{ijk}(A{^{k}}_{b}+ \frac{s\sqrt{\mid\Lambda\mid}}{\gamma}e{^{k}}_{b})\right] \delta^2(x-y) , \nonumber \\
\{\Phi{_{i}}^{a} (x),\Phi{_{j}}^{b}(y) \} &=& -2\frac{1}{\gamma\sqrt{\mid\Lambda\mid}}\epsilon^{0ab}\delta_{ij}\delta^2(x-y),\nonumber\\
\{\Phi{_{i}}^{a} (x),\psi_{j}(y) \} &=&-2\epsilon^{0ab}\left[\delta_{ij}\partial_{b}-\epsilon_{ijk}(A{^{k}}_{b}+ \frac{s\sqrt{\mid\Lambda\mid}}{\gamma}e{^{k}}_{b})\right] \delta^2(x-y),\nonumber \\
\{\Phi{_{i}}^{a} (x),\Psi_{j}(y) \} &=&-2\epsilon^{0ab}\left[\frac{1}{\gamma\sqrt{\mid\Lambda\mid}}\delta_{ij}\partial_{b}-\epsilon_{ijk}(\frac{1}{\gamma\sqrt{\mid\Lambda\mid}}A{^{k}}_{b}+e{^{k}}_{b})\right]\delta^{2}(x-y).\nonumber \\
\{\psi_{i} (x),\psi_{j}(y) \} &=& 0,\nonumber\\
\{\Psi_{i} (x),\Psi_{j}(y) \} &=& 0,\nonumber \\
\{\psi_{i} (x),\Psi_{j}(y) \} &=& 0,
\label{eq17}
\end{eqnarray}
This matrix has a rank=12 and 12 null vectors, thus, the  theory presents a set of 12 first class constraints and 12 second class constraints.   By using the contraction of the null vectors with the constraints (\ref{eq10}) and (\ref{eq15}), we identify  the following 12 first class constraints
\begin{eqnarray}
\gamma^{0}{_{i}}&=&   \pi{_{i}}^{0} \approx 0,  \nonumber \\
\Gamma^{0}{_{i}}&=&   \Pi{_{i}}^{0}  \approx 0,  \nonumber \\
\omega_{i}&=& D_a\chi{_i}^a-s\mid\Lambda\mid \epsilon{_{i}}{^{j}}_{k}e{^{k}}_{a}\Xi{_{j}}^{a}+2\epsilon^{0ab}s\frac{\sqrt{\mid\Lambda\mid}}{\gamma} D_{a}e_{ib}+2\epsilon^{0ab}(F_{iab}+\frac{s\mid\Lambda\mid}{2}\epsilon{_{ijk}}e{^{j}}_{a}e{^{k}}_{b})\approx 0, \nonumber \\
\Gamma_{i}&=&D_a\Xi{_i}^a-\epsilon{_{i}}{^{j}}_{k}e{^{k}}_{a}\chi{_{j}}^{a}+2\epsilon^{0ab}D_{a}e_{ib}+2\epsilon^{0ab}\frac{1}{\gamma\sqrt{\mid\Lambda\mid}}(F_{iab}+\frac{s\mid\Lambda\mid}{2}\epsilon{_{ijk}}e{^{j}}_{a}e{^{k}}_{b}) \approx0,
\label{eq18}
\end{eqnarray}
and the following 12 second class constraints
\begin{eqnarray}
\chi{_{i}}^{a}&=& \pi{_{i}}^{a}- s\frac{\sqrt{\mid\Lambda\mid}}{\gamma}\epsilon^{0ab}\delta_{ij} e{_{b}}^{j} \approx 0,  \nonumber \\
\Xi{_{i}}^{a}&=& \Pi{_{i}}^{a}- 2\epsilon^{0ab}\delta_{ij}(e{_{b}}^{j}+\frac{1}{2\gamma\sqrt{\mid\Lambda\mid}}A{_{b}}^{j})   \approx 0.
\label{eq19}
\end{eqnarray}
It is important to remark  that these constraints have not been reported in the literature, and their  complete structure defined on the full phase space  will be relevant in order to know the fundamental gauge transformations and for  constructing  the Dirac brackets. On the other hand, the constraints will play a key role  to make the  progress in the quantization. We have commented above that the structure of the constraints (\ref{eq18}) is obtained  by means of  the null vectors, for instance,   a set of  null vectors of the  matrix (\ref{eq17}) are  given by
 \begin{eqnarray*}
 V^{1}_i=(0,-\delta^{i}_{j}\partial_{a}\delta^{2}(x-y)-\epsilon^{i}_{jk}A^{k}_{a}\delta^{2}(x-y),0,-s\mid\Lambda\mid\epsilon_{i}{^{j}}_{k}e^{k}_{a}\delta^{2}(x-y),\delta^{i}_{j}\delta^{2}(x-y),0), \nonumber
 \end{eqnarray*}
hence, by performing the contraction of these null vectors with the primary and secondary constraints,   the first class constraint $\omega_i$ given in (\ref{eq18}) is obtained.  \\
Now, we will calculate the algebra of the constraints
\begin{eqnarray}
\{\chi{_{i}}^{a} (x),\chi{_{j}}^{b} (y) \} &=& -2s\frac{\sqrt{\mid\Lambda\mid}}{\gamma}\epsilon^{0ab}\delta_{ij}\delta^{2}(x-y), \nonumber \\
\{\chi{_{i}}^{a} (x),\Xi{_{j}}^{b} (y) \} &=& -2\epsilon^{0ab}\delta_{ij}\delta^{2}(x-y), \nonumber \\
\{ \Xi{_{i}}^{a} (x), \Xi{_{j}}^{b}(y) \} &=& -2\frac{1}{\gamma\sqrt{\mid\Lambda\mid}}\epsilon^{0ab}\delta_{ij}\delta^2(x-y),\nonumber \\
\{\chi{_{i}}^{a} (x),\omega_{j} (y) \} &=&s\mid\Lambda\mid\epsilon{_{ij}}^{k}\Phi{^{a}}_{k}\delta^{2}(x-y)\approx0,\nonumber \\
\{\Xi{_{i}}^{a} (x),\omega_{j} (y) \} &=&\epsilon{_{ij}}^{k}\phi{^{a}}_{k}\delta^{2}(x-y)\approx0, \nonumber \\
\{\chi{_{i}}^{a} (x),\Gamma {_{j}}(y) \} &=&\epsilon{_{ij}}\phi^{a}_{k}\delta^2(x-y)\approx0 , \nonumber \\
\{ \Xi{_{i}}^{a} (x), \Gamma_{j}(y) \} &=& \epsilon_{ijk}\Phi^{ka}\delta^{2}(x-y)\approx0, \nonumber  \\
\{\omega_{i} (x),\omega_{j}(y) \} &=& s\mid\Lambda\mid\epsilon_{ijk}\Gamma^{k}\delta^{2}(x-y)\approx0, \nonumber \\
\{\Gamma{_{i}} (x),\Gamma_{j}(y) \} &=&\epsilon_{ijk}\Gamma^{k} \delta^2(x-y)\approx0 , \nonumber \\
\{ \omega_{i} (x), \Gamma_{j}(y) \} &=& \epsilon_{ijk}\omega^{k}\delta^2(x-y)\approx0.
\label{eq20}
\end{eqnarray}
where we can observe that the algebra is closed. Furthermore, with all the information obtained until  now, we can construct the Dirac brackets. In fact, there are two options for constructing them, the first way is by eliminating the second  class constraints and keeping on  the first class, the second option is by fixing the gauge and converting the first class constraints into  second class ones. In this section we will eliminate the  second class constraints  remaining the firs class ones; in the [FJ] approach we will discuss both. Hence, in order to construct the Dirac brackets, we shall use  the matrix whose elements are only the Poisson brackets among second class constraints, namely $C{_{\alpha\beta}}(u,v)=\{\zeta^{\alpha}(u),\zeta^{\beta}(v)\}$,   given by
\begin{eqnarray}
[C_{(\alpha\beta)}(x,x')]{^{ab}}_{ij}=
-2\left(
 \begin{array}{cccc}
   \frac{s\sqrt{\mid\Lambda\mid}}{\gamma}&1\\
   1&\frac{1}{\gamma\sqrt{\mid\Lambda\mid}}\\
  \end{array}
\right)\delta_{ij}\epsilon^{0ab}\delta^2(x-x'),
\end{eqnarray}
its inverse is given by
\begin{eqnarray}
[C{^{-1}}_{(\alpha\beta)}(x,x')]{^{}}_{}=
\frac{\gamma^{2}}{2(s-\gamma^{2})}\left(
 \begin{array}{cc}
   \frac{1}{\gamma\sqrt{\mid\Lambda\mid}}&-1\\
   -1&\frac{s\sqrt{\mid\Lambda\mid}}{\gamma}\\
  \end{array}
\right)\delta^{ij}\epsilon{_{0ba}}\delta^2(x-x').
\label{eq25a}
\end{eqnarray}
The Dirac  brackets among two functionals $A$, $B$ are expressed by
\begin{eqnarray}
\{A(x),B(y)\}_{D}=\{A(x),B(y)\}_{P}-\int dudv\{A(x),\zeta^{\alpha}(u)\}C{^{-1}}_{\alpha\beta}(u,v)\{\zeta^{\beta}(v),B(y)\},
\label{eq26a}
\end{eqnarray}
where $\{A(x),B(y)\}_{P}$ is the usual Poisson bracket between the functionals $A$, $B$ and $\zeta^{\alpha}(u)=(\chi{_{i}}^{a},\Xi{_{i}}^{a})$ is the set of second class constraints. Hence, by using (\ref{eq25a}) and (\ref{eq26a}) we obtain the following Dirac's brackets of the theory
\begin{eqnarray}
\{e{^{i}}_{a}(x),\pi{^{b}}_{j}(y)\}_{D}&=&\frac{s}{2(s-\gamma^{2})}\delta{^{b}}_{a}\delta{^{i}}_{j}\delta^{2}(x-y),\nonumber \\
\{e{^{i}}_{a}(x),e{^{j}}_{b}(y)\}_{D}&=&\frac{1}{2\sqrt{\mid\Lambda\mid}}\frac{\gamma}{s-\gamma^{2}}\delta{^{ij}}\epsilon_{0ab}\delta^{2}(x-y),\nonumber \\
\{\pi{^{a}}_{i}(x),\pi{^{b}}_{j}(y)\}_{D}&=&\frac{s^{2}}{2\gamma}\frac{\sqrt{\mid\Lambda\mid}}{s-\gamma^{2}}\delta_{ij}\epsilon{^{0ab}}\delta^{2}(x-y),\nonumber \\
\{A{^{i}}_{a}(x),\Pi{^{b}}_{j}(y)\}_{D}&=&\frac{s-2\gamma^{2}}{2(s-\gamma^{2})}\delta{^{b}}_{a}\delta{^{i}}_{j}\delta^{2}(x-y),\nonumber \\
\{A{^{i}}_{a}(x),A{^{j}}_{b}(y)\}_{D}&=&\frac{s\sqrt{\mid\Lambda\mid}}{2}\frac{\gamma}{s-\gamma^{2}}\delta^{ij}\epsilon{_{0ab}}\delta^{2}(x-y), \nonumber \\
\{\Pi{^{a}}_{i}(x),\Pi{^{b}}_{j}(y)\}_{D}&=&\frac{s}{2\gamma\sqrt{\mid\Lambda\mid}}\frac{1}{s-\gamma^{2}}\delta_{ij}\epsilon{^{0ab}}\delta^{2}(x-y),\nonumber \\
\{A{^{i}}_{a}(x),e{^{j}}_{b}(y)\}_{D}&=&\frac{1}{2}\frac{\gamma^{2}}{\gamma^{2}-s}\delta^{ij}\epsilon_{0ab}\delta^{2}(x-y),\nonumber \\
\{e{^{i}}_{a}(x),\Pi{^{b}}_{j}(y)\}_{D}&=&\frac{1}{2\sqrt{\mid\Lambda\mid}}\frac{\gamma}{s-\gamma^{2}}\delta{^{b}}_{a}\delta{^{i}}_{j}\delta^{2}(x-y), \nonumber \\
\{A{^{i}}_{a}(x),\pi{^{b}}_{j}(y)\}_{D}&=&-\frac{s\sqrt{\mid\Lambda\mid}}{2}\frac{\gamma}{s-\gamma^{2}}\delta{^{b}}_{a}\delta{^{i}}_{j}\delta^{2}(x-y), \nonumber \\
\{\pi{^{a}}_{i}(x),\Pi{^{b}}_{j}(y)\}_{D}&=&\frac{1}{2}\frac{s}{s-\gamma^{2}}\delta_{ij}\epsilon{^{0ab}}\delta^{2}(x-y).
\label{eq30}
\end{eqnarray}
We can observe that the Dirac brackets depend of the constants ($s$,  $\gamma$) and we can  reproduce  several scenarios depending of the values  of  these constants.  In fact,  if we take $s=1$ and  the limit $\gamma \rightarrow \infty$ we recover the Dirac  canonical structure of Palatini's  action, for instance, that reported in   \cite{9}. It is important to remark that in [BL] model  the fields $e$, $A$ and its  canonical momenta have become non-commutative while in Palatini's  action they are commutative, this is a classical difference between  [BL] and Palatini's theory. Moreover,  at quantum level  this difference is fundamental for constructing the Ashtekar representation of [BL] model  \cite{4}. \\
Now, we can calculate the gauge transformations on the full phase space.  In fact, the correct gauge symmetry is  obtained  according to Dirac's conjecture  by constructing a gauge generator  using the first class constraints, and the structure of the constraints  defined on the full phase space will give us the fundamental gauge transformations.  For this aim, we will apply the Castellani's algorithm  to construct the gauge generator;
we define the generator of gauge transformations as
\begin{eqnarray}
G=\int_{\sum}\left[D{_{0}}\varepsilon{^{i}}_{0}\gamma{^{0}}_{i}+D{_{0}}\tau{_{0}}^{i}\Gamma{^{0}}_{i}+\varepsilon^{i}\omega{_{i}}+\tau^{i}\Gamma{_{i}}\right].
\end{eqnarray}
Therefore, we find that the gauge transformations on the phase space are

\begin{eqnarray}
\delta{_{0}} e{^{i}}_{0}&=&D_{0}\varepsilon{^{i}}_{0},\nonumber \\
\delta{_{0}} e{^{i}}_{a}&=&-D{_{a}}\varepsilon^{i}+\epsilon{^{i}}_{jk}e{_{a}}^{k}\tau^{j},\nonumber\\
\delta{_{0}} A_{0}{^{i}}&=&D_{0}\tau{_{0}}{^{i}},\nonumber\\
\delta{_{0}} A_{a}{^{i}}&=&-D{_{a}}\tau^{i}+s\mid\Lambda\mid\epsilon{^{i}}_{jk}e{^{k}}_{a}\varepsilon^{j},\nonumber\\
\delta{_{0}} \pi{^{0}}_{i}&=&0 ,\nonumber \\
\delta{_{0}} \pi{^{a}}_{i}&=&-\Omega \epsilon^{0ab}D_{b}\varepsilon_{i} + \epsilon{_{ij}}^{k}\pi{^{a}}_{k}\tau^{j}+s\mid\Lambda\mid\epsilon{_{ij}}^{k}\Xi{^{a}}_{k} ,\nonumber\\
\delta{_{0}} \Pi{^{0}}_{i}&=&-\epsilon{_{ij}}^{k}(\pi{^{0}}_{k}\varepsilon^{j}-\Pi{^{0}}_{k}\tau^{j}),\nonumber\\
\delta{_{0}}\Pi{^{a}}_{i}&=&-2\epsilon^{0ab}D_{b}\varepsilon_{i}+\epsilon{_{ij}}^{k}\chi{^{a}}_{k}\varepsilon^{j}-\frac{1}{\gamma\sqrt{\mid\Lambda\mid}}\epsilon{^{0ab}}D_{b}\tau_{i}+\epsilon{_{ij}}^{k}\Xi{^{a}}_{k}\tau^{j}\nonumber\\ &\,&+2\epsilon^{0ab}\epsilon_{ijk}e{^{k}}_{b}\tau^{j}+\Omega\epsilon^{0ab}\epsilon_{ijk}e^{k}_{b}\varepsilon^{j}.
\label{eq32}
\end{eqnarray}
We realize  that the fundamental  gauge transformations of the [BL] action are given by (\ref{eq32}) and they  are an $\Lambda$-deformed $ISO(3)$ transformations. It is important to  remark, that the internal group of the theory  is $SU(2)$ ( or $SO(3)$), however,  the fundamental gauge symmetry and they correspond to  $\Lambda$-deformed $ISO(3)$ transformations,   all these results were not reported in \cite{4, 5}. On the other hand,  any theory with  background independence is diffeomorphisms covariant, and this symmetry must be obtained from the fundamental gauge transformations. Hence, the diffeomorphisms can be found  by  redefining the gauge parameters as $\varepsilon{_{0}}^{i}=-\varepsilon^{i}=\xi^{\rho}e{^{i}}_{\rho}$, $\tau{_{0}}^{i}=-\tau^{i}=\xi^{\rho}A_{\rho}{^{i}}$, and the gauge transformation (\ref{eq32}) takes the following form
\begin{eqnarray}
e'{^{i}}{_{\alpha}}&\rightarrow&e{^{i}}{_{\alpha}}+\mathfrak{L}{_{\xi}}e{^{i}}{_{\alpha}}+\xi^{\rho}\left[D{_{\alpha}}e{^{i}}{_{\rho}}-D{_{\rho}}e{^{i}}{_{\alpha}}\right] ,  \nonumber \\
A'{_{\alpha}}^{i}&\rightarrow&A{_{\alpha}}^{i}+\mathfrak{L}{_{\xi}}A{_{\alpha}}^{i}+\xi^{\rho}\left[\partial_{\alpha}A{^{i}_{\rho}}-\partial_{\rho}A{^{i}}_{\alpha}+\epsilon_{ijk}A{^{j}}_{\alpha}A{^{k}}_{\rho}+s\mid\Lambda\mid\epsilon{^{i}}_{jk}e{^{j}}_{\alpha}e{^{k}}_{\rho}\right] ,
\label{eq33}
\end{eqnarray}
Therefore, diffeomorphisms are obtained (on shell) from the fundamental gauge transformations as an internal symmetry of the theory. With the correct identification of the constraints, we can carry out the counting of degrees of freedom in the following form: there are $36$ canonical variables $(e{^i}_{\alpha}, A_{\alpha} {^{i}}, \pi^{\alpha}{_{i}},\Pi^{\alpha}{_{i}})$, $12$ first class constraints $(\gamma{_{i}}^{0}, \Gamma{_{i}}^{0}, \omega_{i}, \Gamma_{i})$ and $12$ second class constraints $(\chi{_{i}}^{a}, \Xi{_{i}}^{a})$  and  one  concludes that the  $S_{\gamma}[A,e]$ action for gravity  in three   dimensions is devoid of degrees of freedom, therefore, the theory is topological.\\
As a conclusion of this part,  we have obtained the extended action, the extended Hamiltonian, the complete structure of the constraints on the full phase space, and the algebra among them. The price to pay for working on the complete phase space,  is that  the  theory presents a set of first and second class constraints;  by using the second class constraints we have constructed  the Dirac brackets and they will be useful in the quantization of the theory \cite{4}.\\
\newline
\section{Relation with Chern-Simons Theory }
We have seen in previous sections  that either Palatini's theory  or   exotic action for gravity can be expressed as a Chern-Simons theory, however, will be interesting to  express the BL action as a Chern-Simons theory  as well;  is it possible? the answer  is yes. In fact,    the action analized in the previous section,  can be written in an elegant form in terms of a Chern-Simons theory. By introducing the following  variables $\omega^{\pm i}= A^i \pm \sigma \sqrt{|\Lambda|} e^i$ \cite{5}, where $\sigma^2=s$ with $\sigma =1$  for $\Lambda >0$ and $\sigma= i$ for $\Lambda<0$,  we obtain that the action (\ref{eq1a}) can be written as
\begin{equation}
S{_{\gamma}}= \frac{\sqrt{|\Lambda|}}{2} \left( \gamma ^{-1} + \frac{\sigma}{s}\right)  S_{CS} (\omega^{+})   +\frac{\sqrt{|\Lambda|}}{2} \left( \gamma ^{-1} - \frac{\sigma}{s} \right) S_{CS} (\omega^{-}),
\label{eq3.1}
\end{equation}
where
\begin{equation}
S_{CS}(\omega^{\pm})= \int_M \omega^{\pm i}\wedge d \omega^{\pm}_{i} + \frac{1}{3} \epsilon_{ijk} \omega^{\pm i} \wedge  \omega^{\pm j}  \wedge \omega^{\pm k}. \nonumber
\end{equation}
The equations of motion obtained from  (\ref{eq3.1}) imply that the connections $\omega^{\pm i }$ are flat,  and it is easy to prove  that these  flatness conditions are equivalent to  the equations of motion given in (\ref{eq4})-(\ref{eq5}). On the other hand, if we develop  a pure Dirac's analysis of the action (\ref{eq3.1}) we will reproduce the results given in the previous section, in particular we will reproduce the Dirac brackets. In fact, in summary by performing a pure Dirac's method  we obtain the following results: \\
There are the following first class constraints
\begin{eqnarray}
\mathfrak{G}^{\pm}_{i}=\frac{s\pm\sigma\gamma}{s\gamma\sqrt{\mid\Lambda\mid}}\epsilon^{0ab}(\partial_{a}\omega^{(\pm)}_{bi}+\frac{1}{2}\epsilon_{ijk}\omega^{(\pm)j}_{b}\omega^{(\pm)k}_{c})+D^{(\pm)}_a\chi^{(\pm)a}_{i}\approx 0,
\label{eqxz}
\end{eqnarray}
and there are the  following second class constraints \\
\begin{eqnarray}
\chi^{(\pm)a}_{i}=\pi^{(\pm)a}_{i}-\frac{1}{2}\frac{s \pm \sigma \gamma}{s\gamma\sqrt{\mid\Lambda\mid}}\epsilon^{0ab}\omega_{(\pm)bi}\approx 0,
\label{eqwww}
\end{eqnarray}
here,  $\pi^{(\pm)a}_{i}$ is the conjugate canonical momenta of the connection $\omega^{\pm i}_a$ and $D_a^{\pm} \lambda^i = \partial_a \lambda^i+ \epsilon^{i}{_{jk}} \omega^{\pm j}_a\lambda^k $. Furthermore, by using the second class constraints (\ref{eqwww}) the following Dirac's brackets are obtained
\begin{eqnarray}
\{\omega^{(\pm)i}_{a}(x),\omega^{(\pm)j}_{b}(y)\}_{D}&=& \frac{s\gamma\sqrt{\mid\Lambda\mid}}{s \pm \sigma\gamma}\delta^{ij}\epsilon_{0ab}\delta^{2}(x-y), \nonumber \\
\{\omega{^{(\pm)i}}_{a}(x),\pi^{(\pm)b}_{j}(y)\}_{D}&=&\frac{1}{2}\delta^{i}_{j}\delta^{b}_{a}\delta^{2}(x-y),\nonumber \\
\{\pi^{(\pm)a}_{i}(x),\pi^{(\pm)b}_{j}(y)\}_{D}&=&\frac{1}{4}\frac{s \pm\sigma\gamma}{s\gamma\sqrt{\mid\Lambda\mid}}\delta_{ij}\epsilon{^{0ab}}\delta^{2}(x-y),
\label{eqwwx}
\end{eqnarray}
we can observe  that the Dirac brackets between dynamical variables given in  (\ref{eqwwx}) are depending of  the constants $(s,\gamma,\sigma)$, this fact will be important because by using  the definition of $\omega^{\pm i}_a$ given in terms of $A^i_a$ and $e^i_a$ into (\ref{eqwwx}), then  the Dirac brackets given in (\ref{eq30}) are reproduced. It is important to comment that our results are given in a general form and  contain the cases for $\Lambda >0$ and $\Lambda<0$,  thus,  in particular we can reproduce the results given in \cite{14a} where the case $\Lambda <0$ was studied. On the other hand,  in the limit $\gamma \rightarrow \propto$ the action (\ref{eq3.1}) is reduced to $S_{CS}(\omega)$  and the Dirac brackets are reduced to those reported in \cite{9} where  Palatini's  theory was analyzed. Finally, we can observe that in this section we have proved the equivalence between  [BL] model and the Chern-Simons theory, thus, the standard quantization procedure can be performed. In the following section we will study the action (\ref{eq1a}) by using the [FJ] approach and we will obtain all the Dirac results in an alternative way.
\section{FADDEEV-JACKIW ANALYSIS FOR  BL THEORY }

In this section we will develop the [FJ] formalism for the [BL] model, rewriting  the action (\ref{eq6})  in the following form
\begin{equation}
\mathcal{L}=2\epsilon^{0ab}\delta_{ij}e^{i}_{b}\dot{A}^{j}_{a}+\beta\epsilon^{0ab}\delta_{ij}A^{i}_{b}\dot{A}^{j}_{a}+\Omega\epsilon^{0ab}\delta_{ij}e^{i}_{b}\dot{e}^{j}_{a}-V^{(0)},
\label{eq34}
\end{equation}
where $V^{(0)}=-2\epsilon^{0ab}\delta_{ij}\left[(e^{i}_{0}+\beta A^{i}_{0})(F^{j}_{ab}+s\frac{\mid\Lambda\mid}{2}\epsilon{^{j}}_{kl}e^{k}_{a}e^{l}_{b})+(\Omega e^{i}_{0}+A^{i}_{0})D_{a}e^{j}_{b}\right]$ is called the symplectic potential  and we have introduced the following constants $\Omega$ and $\beta$ defined by
\begin{equation}
\Omega=\frac{s\sqrt{\mid\Lambda\mid}}{\gamma}, \qquad \beta=\frac{1}{\gamma\sqrt{\mid\Lambda\mid}} .
\label{eq35}
\end{equation}
In the [FJ] framework, the Euler-Lagrange equations of motion  are given by  \cite{11}
\begin{equation}
f^{(0)}_{ab}\dot{\xi}^{b}=\frac{\partial V^{(0)}(\xi)}{\partial\xi^{a}},
\label{eq36}
\end{equation}
where the symplectic matrix $f^{(0)}_{ab}$ takes the form
\begin{equation}
f^{(0)}_{ab}(x,y)=\frac{\delta \mathrm{a}_{b}(y)}{\delta\xi^{a}(x)}-\frac{\delta \mathrm{a}_{a}(x)}{\delta\xi^{b}(y)},
\label{37}
\end{equation}
with $\xi{^{(0)a}}$ and $\mathrm{a}{^{(0)}}{_{a}}$ representing  a set of symplectic variables. It is important to comment, that in [FJ] framework we are free to choose  the symplectic variables; we can choose the field configuration variables or  the phase space variables. In previous sections, we have constructed the Dirac brackets  by eliminating the second class constraints, hence,  in order to  obtain these results by means [FJ]  we will use   the configuration space as symplectic variables \cite{13}. For this aim,   we choose   from the symplectic Lagrangian (\ref{eq34})   the following symplectic variables $ \xi{^{(0)a}}(x)=\{e^{i}_{a},e^{i}_{0},A^{i}_{a},A^{i}_{0}\}$  and the components of the symplectic 1-forms are  $\mathrm{a}{^{(0)}}{_{a}}(x)=\{\Omega\epsilon^{0ab}e_{bi},0,2\epsilon^{0ab}e_{bi}+\beta \epsilon^{0ab} A_{bi},0\}$. Hence, by using our set of symplectic variables,  the symplectic matrix (\ref{37}) takes the form
\begin{eqnarray}
f^{(0)}_{ab}(x,y)=
\left(
 \begin{array}{cccc}
   -2\Omega\epsilon^{0ab}\delta_{ij}&0&-2\epsilon^{0ab}\delta_{ij}&0\\
   0&0&0&0\\
   -2\epsilon^{0ab}\delta_{ij}&0&-2\beta\epsilon^{0ab}\delta_{ij}&0\\
   0&0&0&0\\
  \end{array}
\right)\delta^2(x-y).
\label{eq38}
\end{eqnarray}
The symplectic matrix $f^{(0)}_{ab}$ is of dimension  $[18\times18]$   and it is a singular matrix. In fact, in [FJ] method this means that there are  present constraints. In order to obtain these constraints, we calculate  the zero modes of the symplectic matrix, the modes are given by  $(v_{a}^{(0)})_{1}^{T}=(0,v^{e^{i}_{0}},0,0)$ and $(v_{a}^{(0)})_{2}^{T}=(0,0,0,v^{A^{i}_{0}})$, where $v^{e^{i}_{0}}$ and $v^{A^{i}_{0}}$ are  arbitrary functions. In this manner, by using the zero-modes and the symplectic potential $V^{(0)}$ we obtain

\begin{eqnarray}
\Omega^{(0)}_{i}&=&\int d^{2}x(v^{(0)})^{T}_{a}(x)\frac{\delta}{\delta\xi^{(0)a}(x)}\int d^{2}y V^{(0)}(\xi) \nonumber \\
            &=& -\int d^{2}x v^{e^{i}_{0}}(x)2\epsilon^{0ab}\delta_{ij}[(F{^{j}}_{ab}+s\frac{\mid\Lambda\mid}{2}\epsilon{^{j}}_{kl}e^{k}_{a}e^{l}_{b})+\Omega D_{a}e{^{j}}_{b}]\nonumber  \\
&\rightarrow& -2\epsilon^{0ab}\delta_{ij}\left[(F{^{j}}_{ab}+s\frac{\mid\Lambda\mid}{2}\epsilon{^{j}}_{kl}e^{k}_{a}e^{l}_{b})+\Omega D_{a}e{^{j}}_{b}\right]=0,
    \label{eq39}
\end{eqnarray}
\begin{eqnarray}
\beta^{(0)}_{i}&=&\int d^{2}x(v^{(0)})^{T}_{a}(x)\frac{\delta}{\delta\xi^{(0)a}(x)}\int d^{2}y V^{(0)}(\xi) \nonumber \\
            &=& -\int d^{2}x v^{A^{i}_{0}}(x)2\epsilon^{0ab}\delta_{ij}\left[\beta(F{^{j}}_{ab}+s\frac{\mid\Lambda\mid}{2}\epsilon{^{j}}_{kl}e^{k}_{a}e^{l}_{b})+  D_{a}e{^{j}}_{b}\right] \nonumber  \\
&\rightarrow& -2\epsilon^{0ab}\delta_{ij}\left[\beta(F{^{j}}_{ab}+s\frac{\mid\Lambda\mid}{2}\epsilon{^{j}}_{kl}e^{k}_{a}e^{l}_{b})+  D_{a}e{^{j}}_{b}\right]=0.
          \label{eq40}
\end{eqnarray}
thus we identify the following constraints
\begin{eqnarray}
\Omega^{(0)}_{i}=2\epsilon^{0ab}\delta_{ij}\left[(F{^{j}}_{ab}+s\frac{\mid\Lambda\mid}{2}\epsilon{^{j}}_{kl}e^{k}_{a}e^{l}_{b})+\Omega D_{a}e{^{j}}_{b}\right]=0,
\label{eq38a}
\end{eqnarray}
\begin{eqnarray}
\beta^{(0)}_{i}= 2\epsilon^{0ab}\delta_{ij}\left[\beta(F{^{j}}_{ab}+s\frac{\mid\Lambda\mid}{2}\epsilon{^{j}}_{kl}e^{k}_{a}e^{l}_{b})+  D_{a}e{^{j}}_{b}\right]=0,
\label{eq39a}
\end{eqnarray}
these constraints are the secondary constraints found by means Dirac's method in the above sections. In order to observe  if the are more constraints, we calculate the following  \cite{14, 15, 16, 17}
\begin{eqnarray}
   f^{(1)}_{cb}\dot{\xi}^{b}=Z_{c}(\xi),
\label{eq41}
\end{eqnarray}

where
\begin{eqnarray}
Z_{c}(\xi)=
\left(
 \begin{array}{cccc}
   \frac{\partial V^{(0)}(\xi)}{\partial \xi^{a}}\\
   0\\
   0\\
  \end{array}
\right),
\label{eq42}
\end{eqnarray}
and
\begin{eqnarray}
f^{(1)}_{cb}=
\left(
 \begin{array}{cccc}
   f^{(0)}_{ab}\\
   \frac{\partial\Omega^{(0)}}{\partial\xi^{b}}\\
   \frac{\partial\beta^{(0)}}{\partial\xi^{b}}\\
  \end{array}
\right)=\left(
 \begin{array}{cccc}
   -2\Omega\epsilon^{0ab}\delta_{ij}&0&-2\epsilon^{0ab}\delta_{ij}&0\\
   0&0&0&0\\
   -2\epsilon^{0ab}\delta_{ij}&0&-2\beta\epsilon^{0ab}\delta_{ij}&0\\
   0&0&0&0\\
   2\epsilon^{0ab}(\Omega\delta_{ij}\partial_{a}-\epsilon_{ijk}(\Omega A^{k}_{a}+s\mid\Lambda\mid e^{k}_{a}))&0& 2\epsilon^{0ab}(\delta_{ij}\partial_{a}-\epsilon_{ijk}( A^{k}_{a}+ \Omega e^{k}_{a}))&0\\
    2\epsilon^{0ab}(\delta_{ij}\partial_{a}-\epsilon_{ijk}(A^{k}_{a}+\Omega e^{k}_{a}))&0& 2\epsilon^{0ab}(\beta\delta_{ij}\partial_{a}-\epsilon_{ijk}(\beta A^{k}_{a}+e^{k}_{a}))&0\\
  \end{array}
\right)\delta^2(x-y). \nonumber \\
\label{eq43}
\end{eqnarray}
We can observe that the  matrix (\ref{eq43})  is  not a square matrix as expected, however, it has   linearly independent modes given by $(v^{(1)})_{1}^{T}=(\delta^{i}_{j}\partial_{a}v^{\lambda}+\epsilon{^{i}}_{kj}A^{k}_{a}v^{\lambda},\delta^{i}_{j}v^{e^{i}_{0}},-s\mid\Lambda\mid\epsilon{^{i}}_{jk}e^{k}_{a},0,\delta^{i}_{j}v^{\lambda},0)$ and $(v^{(1)})_{2}^{T}=(-\epsilon{^{i}}_{jk}e^{k}_{a},0,\delta^{i}_{j}\partial_{a}v^{\beta}+\epsilon{^{i}}_{kj}A^{k}_{a}v^{\beta},\delta^{i}_{j}v^{A^{i}_{0}},0,\delta^{i}_{j}v^{\beta})$. These modes are used in order to obtain further  constraints. In fact, by calculating the following contraction \cite{14, 15, 16, 17}
\begin{equation}
(v^{(1)})_{c}^{T}Z_{c}=0,
\label{eq44}
\end{equation}
where $c=1,2$, we obtain that (\ref{eq44}) is an identity, thus, in [FJ]  formalism there are not more constraints for the theory under study. \\
Now, we will construct a new symplectic Lagrangian with   the information of the constraints obtained in (\ref{eq38a}) and (\ref{eq39a}). In order to archive this aim, we introduce  $e^{i}_{0}=\dot{\lambda}^{i}$ and  $A^{i}_{0}=\dot{\theta}^{i}$,  as Lagrange multipliers associated to those constraints, thus, we obtain the following symplectic Lagrangian

\begin{equation}
\mathcal{L}^{(1)}=2\epsilon^{0ab}\delta_{ij}e^{i}_{b}\dot{A}^{j}_{a}+\beta\epsilon^{0ab}\delta_{ij}A^{i}_{b}\dot{A}^{j}_{a}+\Omega\epsilon^{0ab}\delta_{ij}e^{i}_{b}\dot{e}^{j}_{a}+ \Omega^{(0)}_{i}\dot{\lambda}^{i}+\beta^{(0)}_{i}\dot{\theta}^{i}-V^{(1)},
\label{eq45}
\end{equation}

where $V^{(1)}=V^{(0)}\mid_{\Omega^{(0)}_{i}=0,\beta^{(0)}_{i}=0}=0$, the symplectic potential vanishes  reflecting the general covariance of the theory. In this manner, from (\ref{eq45}) we identify the following new symplectic variables $\xi{^{(1)a}}(x)=\{e^{i}_{a},\lambda^{i},A^{i}_{a},\theta^{i}\}$ and the new symplectic 1-forms
$\mathrm{a}{^{(0)}}{_{a}}(x)=\{\Omega\epsilon^{0ab}e_{bi},\Omega^{(0)}_{i},2\epsilon^{0ab}e_{bi}+\beta \epsilon^{0ab} A_{bi},\beta^{(0)}_{i}\}$. Hence, by using the new symplectic variables and 1-forms, we can calculate the following symplectic matrix

\begin{eqnarray}
f^{(1)}_{ab}(x,y)=\nonumber
\end{eqnarray}

\begin{eqnarray*}
{\tiny
\left(
 \begin{array}{cccc}
   -2\delta_{ij}&-2(\Omega\delta_{ij}\partial_{b}+\epsilon_{ijk}(\Omega A^{k}_{b}+s\mid\Lambda\mid e^{k}_{b}))&-2\delta_{ij}&-2(\delta_{ij}\partial_{b}+\epsilon_{ijk}(A^{k}_{b}+\Omega e^{k}_{b}))\\
   -2(\Omega\delta_{ij}\partial_{a}-\epsilon_{ijk}(\Omega A^{k}_{i}+s\mid\Lambda\mid e^{k}_{a} ))&0&-2(\Omega\delta_{ij}\partial_{a}-\epsilon_{ijk}(A^{k}_{i}+\Omega e^{k}_{a}))&0\\
    -2\delta_{ij}&-2(\delta_{ij}\partial_{b}+ \epsilon_{ijk}(A ^{k}_{b}+\Omega e^{k}_{b}))&-2\beta\delta_{ij}&-2(\beta\delta_{ij}\partial_{b}+\epsilon_{ijk}(\beta A^{k}_{b}+e^{k}_{b}))\\
  -2(\delta_{ij}\partial_{a}-\epsilon_{ijk}(A^{k}_{a}+ \Omega e^{k}_{a}))&0&-2(\beta\delta_{ij}\partial_{a}-\epsilon_{ijk}(\beta A^{k}_{a}+e^{k}_{a}))&0\\
  \end{array}
\right)} \nonumber \\
\label{eq46}
\end{eqnarray*}

\begin{eqnarray}
\times\epsilon^{0ab}\delta^2(x-y)
\end{eqnarray}
The symplectic matrix $f^{(1)}_{ab}$ represents a $[18\times18]$ singular matrix. However, we have commented that there are not more constraints; the noninvertibility of (\ref{eq46}) means that the theory has a gauge symmetry. In order to invert the symplectic matrix  we choose the following gauge fixing as constraints
\begin{eqnarray}
A^{i}_{0}(x)&=&0,  \nonumber\\
e^{i}_{0}(x)&=&0, \nonumber
\end{eqnarray}
then we  introduce the   Lagrangians multipliers $\phi_{i}$ and $\alpha_{i}$ associated with the above gauge fixing for constructing  a new  symplectic Lagrangian. Now the symplectic Lagrangian is given by
\begin{equation}
\mathcal{L}^{(2)}=2\epsilon^{0ab}\delta_{ij}e^{i}_{b}\dot{A}^{j}_{a}+\beta\epsilon^{0ab}\delta_{ij}A^{i}_{b}\dot{A}^{j}_{a}+\Omega\epsilon^{0ab}\delta_{ij}e^{i}_{b}\dot{e}^{j}_{a}+ (\Omega^{(0)}_{i}+\phi_{i})\dot{\lambda}^{i}+(\beta^{(0)}_{i}+\alpha_{i})\dot{\theta^{i}},
\label{eq47}
\end{equation}

thus, we identify the following set of symplectic variables $\xi{^{(2)a}}(x)=\{e^{i}_{a},\lambda^{i},\phi_{i},A^{i}_{a},\theta^{i},\alpha_{i}\}$  and the symplectic 1-forms $\mathrm{a}{^{(0)}}{_{a}}(x)=\{\Omega\epsilon^{0ab}e_{bi},\Omega^{(0)}_{i}+\phi_{i},0,2\epsilon^{0ab}e_{bi}+\beta \epsilon^{0ab} A_{bi},\beta^{(0)}_{i}+\alpha_{i},0\}$. Furthermore, by using these symplectic variables we find that the symplectic matrix is given by

\begin{eqnarray}
f^{(2)}_{ab}(x,y)=\nonumber
\end{eqnarray}

\begin{eqnarray*}
{\tiny
\left(
 \begin{array}{cccccc}
   -2\Omega\delta_{ij}&-2(\Omega\delta_{ij}\partial_{b}+\epsilon_{ijk}(\Omega A^{k}_{b}+s\mid\Lambda\mid e^{k}_{b}))&0&-2\delta_{ij}&-2(\delta_{ij}\partial_{b}+\epsilon_{ijk}(A^{k}_{b}+\Omega e^{k}_{b}))&0\\
   -2(\Omega\delta_{ij}\partial_{a}-\epsilon_{ijk}(\Omega A^{k}_{i}+s\mid\Lambda\mid e^{k}_{a} ))&0&-\delta^{j}_{i}&-2(\Omega\delta_{ij}\partial_{a}-\epsilon_{ijk}(A^{k}_{i}+\Omega e^{k}_{a}))&0&0\\
   0&\delta^{i}_{j}&0&0&0&0\\
    -2\delta_{ij}&-2(\delta_{ij}\partial_{b}+ \epsilon_{ijk}(A ^{k}_{b}+\Omega e^{k}_{b}))&0&-2\beta\delta_{ij}&-2(\beta\delta_{ij}\partial_{b}+\epsilon_{ijk}(\beta A^{k}_{b}+e^{k}_{b}))&0\\
  -2(\delta_{ij}\partial_{a}-\epsilon_{ijk}(A^{k}_{a}+ \Omega e^{k}_{a}))&0&0&-2(\beta\delta_{ij}\partial_{a}-\epsilon_{ijk}(\beta A^{k}_{a}+e^{k}_{a}))&0&-\delta^{j}_{i}\\
  0&0&0&0&\delta^{i}_{j}&0\\
  \end{array}
\right)} \nonumber \\
\label{eq48}
\end{eqnarray*}

\begin{eqnarray}
\times\epsilon^{0ab}\delta^2(x-y)
\end{eqnarray}
The symplectic matrix $f^{(2)}_{ab}$ represents a $[24\times24]$ nonsingular matrix, hence, it is a symplectic tensor. After a long calculation, the inverse is given by
\begin{eqnarray}
[f^{(2)}_{ab}(x,y)]^{-1}=\nonumber
\end{eqnarray}
\begin{eqnarray}
{\tiny
\left(
 \begin{array}{cccccc}
   \frac{\gamma}{2\sqrt{\mid\Lambda\mid}(s-\gamma^{2})}\epsilon_{0ab}\delta^{ij}&0&\delta^{i}_{j}\partial_{a}-\epsilon{^{i}}_{jk}A^{k}_{a}&-\frac{\gamma^{2}}{2(s-\gamma^{2})}\epsilon_{0ab}\delta^{ij}&0&-\epsilon{^{i}}_{jk}e^{k}_{a}\\
   0&0&\delta^{j}_{i}&0&0&0\\
\delta^{j}_{i}\partial_{b}-\epsilon{_{i}}{^{j}}_{k}A^{k}_{b}&-\delta^{i}_{j}&0&-s\mid\Lambda\mid\epsilon{_{i}}{^{j}}_{k}e^{k}_{b}&0&0\\
-\frac{\gamma^{2}}{2(s-\gamma^{2})}\epsilon_{0ab}\delta^{ij}&0&-s\mid\Lambda\mid\epsilon{^{i}}_{jk}e^{k}_{a}& \frac{s\gamma\sqrt{\mid\Lambda\mid}}{2(s-\gamma^{2})}\epsilon_{0ab}\delta^{ij}&0&\delta^{i}_{j}\partial_{a}-\epsilon{^{i}}_{jk}A^{k}_{a}\\
0&0&0&0&0&\delta^{j}_{i}\\
-\epsilon{_{i}}{^{j}}_{k}e^{k}_{b}&0&0&\delta^{j}_{i}\partial_{b}-\epsilon{_{i}}{^{j}}_{k}A^{k}_{b}&-\delta^{i}_{j}&0\\
  \end{array}
\right)\delta^2(x-y)}. \nonumber \\
\label{eq49}
\end{eqnarray}
Therefore, from (\ref{eq49}) it is possible to identify the following [FJ] generalized brackets by means of
\begin{eqnarray}
\{\xi_{i}^{(2)}(x),\xi_{j}^{(2)}(y)\}_{FD}=[f^{(2)}_{ij}(x,y)]^{-1},
\end{eqnarray}
thus, the following brackets are identified
\begin{eqnarray}
\{e^{i}_{a}(x),e^{j}_{b}(y)\}_{FD}&=&\frac{\gamma}{2\sqrt{\mid\Lambda\mid}(s-\gamma^{2})}\epsilon_{0ab}\delta^{ij}\delta^{2}(x-y),\nonumber\\
\{A^{i}_{a}(x),e^{j}_{b}(y)\}_{FD}&=&-\frac{\gamma^{2}}{2(s-\gamma^{2})}\epsilon_{0ab}\delta^{ij}\delta^{2}(x-y),\nonumber\\
\{A^{i}_{a}(x),A^{j}_{b}(y)\}_{FD}&=&\frac{s\sqrt{\mid\Lambda\mid}\gamma}{2(s-\gamma^{2})}\epsilon_{0ab}\delta^{ij}\delta^{2}(x-y),\nonumber\\
\{e^{i}_{a}(x),\phi_{j}(y)\}_{FD}&=&(\delta^{i}_{j}\partial_{a}-\epsilon{^{i}}_{jk}A^{k}_{a})\delta^{2}(x-y),\nonumber\\
\{A^{i}_{a}(x),\alpha_{j}(y)\}_{FD}&=&(\delta^{i}_{j}\partial_{a}-\epsilon{^{i}}_{jk}A^{k}_{a})\delta^{2}(x-y),\nonumber\\
\{e^{i}_{a}(x),\alpha_{j}(y)\}_{FD}&=&-\epsilon{^{i}}_{jk}e^{k}_{a},\nonumber\\
\{A^{i}_{a}(x),\phi_{j}(y)\}_{FD}&=&-s\mid\Lambda\mid\epsilon{^{i}}_{jk}e^{k}_{a},\nonumber\\
\{\lambda^{i}(x),\phi_{j}(y)\}_{FD}&=&\delta^{j}_{i}\delta^{2}(x-y),\nonumber\\
\{\theta^{i}(x),\alpha_{j}(y)\}_{FD}&=&\delta^{j}_{i}\delta^{2}(x-y).
\label{eq50}
\end{eqnarray}
It is important to comment,  that the generalized [FJ] brackets coincide with those obtained by means of  the Dirac method reported in the  previous section. In fact, if we perform  a redefinition of the fields introducing the momenta given by
\begin{eqnarray}
\pi{_{i}}^{a}&=&s\frac{\sqrt{\mid\Lambda\mid}}{\gamma}\epsilon^{0ab}\delta_{ij} e{_{b}}^{j},\nonumber \\ \Pi{_{i}}^{a}&=&2\epsilon^{0ab}\delta_{ij}(e{_{b}}^{j}+\frac{1}{2\gamma\sqrt{\mid\Lambda\mid}}A{_{b}}^{j}),
\label{eq51}
\end{eqnarray}
the generalized [FJ] brackets (\ref{eq50}) take the form
\begin{eqnarray}
\{e{^{i}}_{a}(x),\pi{^{b}}_{j}(y)\}_{FD}&=&\frac{s}{2(s-\gamma^{2})}\delta{^{b}}_{a}\delta{^{i}}_{j}\delta^{2}(x-y),\nonumber\\
\{e{^{i}}_{a}(x),e{^{j}}_{b}(y)\}_{FD}&=&\frac{1}{2\sqrt{\mid\Lambda\mid}}\frac{\gamma}{s-\gamma^{2}}\delta{^{ij}}\epsilon_{0ab}\delta^{2}(x-y),\nonumber\\
\{\pi{^{a}}_{i}(x),\pi{^{b}}_{j}(y)\}_{FD}&=&\frac{s^{2}}{2\gamma}\frac{\sqrt{\mid\Lambda\mid}}{s-\gamma^{2}}\delta_{ij}\epsilon{^{0ab}}\delta^{2}(x-y),\nonumber\\
\{A{^{i}}_{a}(x),\Pi{^{b}}_{j}(y)\}_{FD}&=&\frac{s-2\gamma^{2}}{2(s-\gamma^{2})}\delta{^{b}}_{a}\delta{^{i}}_{j}\delta^{2}(x-y),\nonumber\\
\{A{^{i}}_{a}(x),A{^{j}}_{b}(y)\}_{FD}&=&\frac{s\sqrt{\mid\Lambda\mid}}{2}\frac{\gamma}{s-\gamma^{2}}\delta^{ij}\epsilon{_{0ab}}\delta^{2}(x-y),\nonumber\\
\{\Pi{^{a}}_{i}(x),\Pi{^{b}}_{j}(y)\}_{FD}&=&\frac{s}{2\gamma\sqrt{\mid\Lambda\mid}}\frac{1}{s-\gamma^{2}}\delta_{ij}\epsilon{^{0ab}}\delta^{2}(x-y),\nonumber\\
\{A{^{i}}_{a}(x),e{^{j}}_{b}(y)\}_{FD}&=&\frac{1}{2}\frac{\gamma^{2}}{\gamma^{2}-s}\delta^{ij}\epsilon_{0ab}\delta^{2}(x-y),\nonumber\\
\{e{^{i}}_{a}(x),\Pi{^{b}}_{j}(y)\}_{FD}&=&\frac{1}{2\sqrt{\mid\Lambda\mid}}\frac{\gamma}{s-\gamma^{2}}\delta{^{b}}_{a}\delta{^{i}}_{j}\delta^{2}(x-y),\nonumber\\
\{A{^{i}}_{a}(x),\pi{^{b}}_{j}(y)\}_{FD}&=&-\frac{s\sqrt{\mid\Lambda\mid}}{2}\frac{\gamma}{s-\gamma^{2}}\delta{^{b}}_{a}\delta{^{i}}_{j}\delta^{2}(x-y),\nonumber\\
\{\pi{^{a}}_{i}(x),\Pi{^{b}}_{j}(y)\}_{FD}&=&\frac{1}{2}\frac{s}{s-\gamma^{2}}\delta_{ij}\epsilon{^{0ab}}\delta^{2}(x-y),
\end{eqnarray}
where we can observe that coincide with  the full Dirac's brackets found in (\ref{eq30}).  \\
Furthermore, we have commented above that in [FJ] approach it is not necessary classify   the constraints in first class  and  second class, in [FJ] formulation all the constraints are at the same footing. Thus, we can carry out the counting of physical degrees of freedom in the following form; there are $12$  dynamical variables $(e^{i}_{a}, A^{i}_{a})$ and $12$ constraints $(\Omega^{(0)}_{i} , \beta^{(0)}_{i}, A^{i}_{0}, e^{i}_{0} )$, therefore, the theory lacks of physical degrees of freedom.\\
We finish this section by calculating the gauge transformations of the theory, for this aim we calculate the modes of the matrix (\ref{eq46})
\begin{eqnarray}
(w^{(1)})_{1}^{T}=(-\delta^{i}_{j}\partial_{a}\delta^{2}(x-y)-\epsilon{^{i}}_{jk}A^{k}_{a}\delta^{2}(x-y),\delta^{i}_{j}\delta^{2}(x-y),-s\mid\Lambda\mid\epsilon{^{i}}_{jk}e^{k}_{a}\delta^{2}(x-y),0),\nonumber  \\
\label{eq.53s}
\end{eqnarray}
\begin{eqnarray}
(w^{(1)})_{2}^{T}=(-\epsilon{^{i}}_{jk}e^{k}_{a}\delta^{2}(x-y),0,-\delta^{i}_{j}\partial_{a}\delta^{2}(x-y)-\epsilon{^{i}}_{jk}A^{k}_{a}\delta^{2}(x-y),\delta^{i}_{j}\delta^{2}(x-y)).
\label{eq.54s}
\end{eqnarray}
In agreement with the [FJ] symplectic formalism, the zero-modes $(w^{(1)})_{1}^{T}$ and $(w^{(1)})_{2}^{T}$ are the generators of  infinitesimal gauge transformation of the action (\ref{eq34})  and are given by
\begin{eqnarray*}
\delta e^{i}_{a}(x)&=&-D_{a}\varepsilon^{i}+\epsilon{^{i}}_{jk}e^{k}_{a}\tau^{j} ,\nonumber \\
\delta e^{i}_{0}(x)&=&\partial_{0}\varepsilon^{i}, \nonumber \\
\delta A^{i}_{a}(x)&=&-D_{a}\tau^{i}+s\mid\Lambda\mid\epsilon{^{i}}_{jk}e^{k}_{a}\varepsilon^{j}, \nonumber \\
\delta A^{i}_{0}(x)&=&\partial_{0}\tau^{i}.\nonumber
\end{eqnarray*}
In fact, the mode (\ref{eq.53s}) is the generator of translations and  the mode (\ref{eq.54s}) is the generator of rotations.  In this manner, by using the [FJ] symplectic framework we have reproduced the $\Lambda$-deformed $ISO(3)$ gauge transformations  reported  by means of Dirac's method. Finally, in order to complete our work,  in Appendix A  we have developed the Dirac analysis for the Abelian case. In that appendix, we performed  the full constraints program and we  have  constructed  the Dirac brackets by fixing the gauge,  then in Appendix B we reproduce all Dirac's results in a more economical way by using [FJ] framework.
\section{ Conclusions and prospects}
In this paper a pure Dirac's formalism and a full [FJ] approach for [BL] model have been performed. With respect to Dirac's method,  the complete structure of the constraints was found and the algebra between the constraints defined on the full phase space was developed. Furthermore, we observed that the internal group is $SU(2)$ (or SO(3)), however,  the fundamental gauge symmetry  correspond to an $\Lambda$-deformed $ISO(3)$ transformations. In addition, we have eliminated the second class constraints by introducing the Dirac brackets, and   we observed that the Dirac brackets depends on the $\gamma$ parameter and this fact makes classically different  [BL]  from Palatini's  theory. On the other hand, we have reproduced  all the relevant Dirac's results by performing the [FJ] framework. In fact, we have found the  [FJ] constraints  and we have showed that there are less constraints than those found with Dirac's method. Moreover, we have showed that the generalized [FJ] brackets and the Dirac's ones coincide to each other. In this manner, we have reproduced all relevant Dirac's results by working with  [FJ], in particular we can see that [FJ] is more economical than Dirac's method. In addition, we proved the equivalence between [BL] model and Chern-Simons theory;  from a pure Dirac's analysis of the  Chern-Simons theory,   all  relevant Dirac's results obtained   using the connection and the frame fields  as dynamical variables were reproduced. Finally, we would to comment that our [FJ] analysis is  generic and we can reproduce all the results reported in \cite{4} where Dirac's method was employed. In fact, it is straightforward  observe that using the Axial gauge in the matrix (\ref{eq46}),    the [BL] theory can be written in terms of a  $SO(3)$ Ashtekar connection. This result is obtained  in more economical way  by using the [FJ] framework. Hence, our results extend   those results found in the literature. \\
\newline
\newline
\newline
\noindent \textbf{Acknowledgements}\\[1ex]
This work was supported by CONACyT under Grant No. CB-2014-01/ 240781. We would like to
thank R. Cartas-Fuentevilla for discussion on the subject and reading of the manuscript.

\appendix
\section{Canonical analysis of the [BL] Abelian theory}
In this appendix, we shall resume the canonical analysis of the Abelian version of [BL] action given by
\begin{eqnarray}
S^{Abelian}[A,e]&=&\int2e^{i}\wedge F_{i}[A]+\frac{1}{\sqrt{\mid \Lambda\mid}}\int A^{i}\wedge d A_{i}+s\sqrt{\mid\Lambda\mid} e^{i}\wedge de_{i} \nonumber \\
                &=&\int\epsilon^{0ab}\Bigg[\left(\frac{A^{i}_{0}}{\gamma\sqrt{\mid\Lambda\mid}}+e^{i}_{0}\right)F_{abi}+\left(\frac{A^{i}_{b}}{\gamma\sqrt{\mid\Lambda\mid}}+e^{i}_{b}\right)\dot{A}_{ai}+\left(\frac{s\sqrt{\mid\Lambda\mid}}{\gamma}e^{i}_{0}+A^{i}_{0}\right)T_{abi}\nonumber\\ &\,&+\left(\frac{s\sqrt{\mid\Lambda\mid}}{\gamma}e^{i}_{b}+A^{i}_{b}\right)\dot{e}_{ai}\Bigg],
\label{eq122}
\end{eqnarray}
where $A_{\mu}^{a}$ and $e^{a}_\mu$ are a set of three $U(1)$ vector fields, $F_{ab}^{i}=\partial_a A^i_b-\partial_b A^i_a$, $T_{ab}= \partial_a e^i_b-\partial_b e^i_a$.  By introducing the canonical momenta  defined by
\begin{equation}
\pi^{\lambda}_{i}:=\frac{\partial\mathcal{L}}{\partial\dot{A}^{i}_{\lambda}}=\epsilon^{0\lambda\rho}\left[\frac{1}{\sqrt{\mid\Lambda\mid}\gamma}A_{\rho i}+e_{\rho i}\right],
\label{eq125}
\end{equation}
\begin{equation}
p^{\lambda}_{i}:=\frac{\partial\mathcal{L}}{\partial\dot{e}^{i}_{\lambda}}=\epsilon^{0\lambda\rho}\left[A_{\rho i}+\frac{s\sqrt{\mid\Lambda\mid}}{\gamma} e_{\rho i}\right].
\label{eq126}
\end{equation}
and performing the canonical analysis, we obtain the following results: we find  4 fist class constraints
\begin{eqnarray}
\gamma^{1}&=& p^{0}_{i} \approx 0,  \nonumber \\
\gamma^{2}&=& 2\partial_{a}p^{a}_{i}-\partial_{a}\phi^{a}_{i} \approx 0, \nonumber \\
\gamma^{3}&=& \pi^{0}_{i} \approx 0, \nonumber \\
\gamma^{4}&=& 2\partial_{a}\pi^{a}_{i}-\partial_{a}\Phi^{a}_{i} \approx 0,
\label{eq135}
\end{eqnarray}
and the following 4 second class constraints
\begin{eqnarray}
\chi^{a}_{1i}&=& p^{a}_{i}-\epsilon^{0ab}\left[A_{b i}+ \Omega e_{b i}\right]\approx 0,\nonumber \\
\chi^{a}_{2i}&=& \pi^{a}_{i}-\epsilon^{0ab}\left[\beta A_{b i}+e_{b i}\right] \approx0.
\label{eq136}
\end{eqnarray}
Now, the nontrivial algebra between the constraints is given by the algebra of the second class constraints as expected
\begin{eqnarray}
\{\chi^{a}_{1i}(x),\chi^{b}_{1j}(y) \} &=&-2\Omega\epsilon^{0ab}\delta_{ij}\delta^{2}(x-y), \nonumber \\
\{\chi^{a}_{1i}(x),\chi^{b}_{2j}(y) \} &=&-2\epsilon^{0ab}\delta_{ij}\delta^{2}(x-y), \nonumber \\
\{\chi^{a}_{2i} (x),\chi^{b}_{2j}(y) \} &=&-2\beta\epsilon^{0ab}\delta_{ij}\delta^{2}(x-y).
\label{eq138}
\end{eqnarray}
In order to construct the Dirac brackets by eliminating the second class constraints, we  write in matrix form the Poisson brackets among second class constraints, namely
\begin{eqnarray}
C^{ab}=
\left(
 \begin{array}{cccc}
   -2\Omega&-2\\
   -2&-2\beta\\
  \end{array}
\right)\epsilon^{0ab}\delta_{ij}\delta^2(x-y),
\end{eqnarray}
and we calculate its inverse given by
\begin{eqnarray}
[C^{ab}]{^{-1}}=\frac{\gamma^{2}}{2(s-\gamma^{2})}
\left(
 \begin{array}{cccc}
   \beta&-1\\
   -1&\Omega\\
  \end{array}
\right)\epsilon_{0ab}\delta^{ij}\delta^2(x-y).
\label{eq61a}
\end{eqnarray}
Hence,  by using the matrix (\ref{eq61a}) we obtain the following Dirac's brackets of the theory
\begin{eqnarray}
\{e^{i}_{a}(x),e^{j}_{b}(y)\}_{D}&=&\frac{\gamma}{2\sqrt{\mid\Lambda\mid}(s-\gamma^{2})}\epsilon_{0ab}\delta^{ij}\delta^{2}(x-y), \nonumber\\
\{A^{i}_{a}(x),e^{j}_{b}(y)\}_{D}&=&-\frac{\gamma^{2}}{2(s-\gamma^{2})}\epsilon_{0ab}\delta^{ij}\delta^{2}(x-y),\nonumber\\
\{e^{i}_{a}(x),p^{b}_{j}(y)\}_{D}&=&\frac{1}{2}\delta^{b}_{a}\delta^{i}_{j}\delta^{2}(x-y),\nonumber\\
\{p^{a}_{i}(x),p^{b}_{j}(y)\}_{D}&=&\frac{s\sqrt{\mid\Lambda\mid}}{2\gamma}\epsilon^{0ab}\delta_{ij}\delta^{2}(x-y),\nonumber\\
\{A^{i}_{a}(x),A^{j}_{b}(y)\}_{D}&=&\frac{s\sqrt{\mid\Lambda\mid}\gamma}{2(s-\gamma^{2})}\epsilon_{0ab}\delta^{ij}\delta^{2}(x-y),\nonumber\\
\{A^{i}_{a}(x),\pi^{b}_{j}(y)\}_{D}&=&\frac{1}{2}\delta^{b}_{a}\delta^{i}_{j}\delta^{2}(x-y), \nonumber\\
\{\pi^{a}_{i}(x),\pi^{b}_{j}(y)\}_{D}&=&\frac{\beta}{2}\epsilon^{0ab}\delta_{ij}\delta^{2}(x-y), \nonumber\\
\{p^{a}_{i}(x),\pi^{b}_{j}(y)\}_{D}&=&\frac{1}{2}\epsilon^{0ab}\delta_{ij}\delta^{2}(x-y), \nonumber \\
\{e^{i}_{0}(x),p^{0}_{j}(y)\}_{D}&=&\delta^{i}_{j}\delta^{2}(x-y),\nonumber\\
\{A^{i}_{0}(x),\pi^{0}_{j}(y)\}_{D}&=&\delta^{i}_{j}\delta^{2}(x-y),
\label{eq43}
\end{eqnarray}
hence,  the Dirac brackets for Abelian and non-Abelian theory  coincide to each other. In the following  lines, we will construct the Dirac brackets by fixing the gauge,  then we will reproduce these results by means [FJ] framework.
\subsection{Dirac's brackets by fixing the gauge}
In order  to construct the Dirac brackets by fixing the gauge,  it is necessary to convert the first class constraints into second class, we will work  with the temporal and Coulomb gauge
\begin{eqnarray}
\Omega_{1}&=&e^{i}_{0}\approx0, \nonumber \\
\Omega_{2}&=&\partial^{a}e^{i}_{a}\approx0, \nonumber \\
\Omega_{3}&=&A^{i}_{0}\approx0, \nonumber \\
\Omega_{4}&=&\partial^{a}A^{i}_{a}\approx0, \nonumber \\
\Omega_{5}&=& p^{0}_{i} \approx 0,  \nonumber \\
\Omega_{6}&=& 2\partial_{a}p^{a}_{i}-\partial_{a}\chi^{a}_{1i} \approx 0, \nonumber \\
\Omega_{7}&=& \pi^{0}_{i} \approx 0, \nonumber \\
\Omega_{8}&=& 2\partial_{a}\pi^{a}_{i}-\partial_{a}\chi^{a}_{2i} \approx 0, \nonumber \\
\Omega_{9}&=& p^{a}_{i}-\epsilon^{0ab}\left[A_{b i}+ \Omega e_{b i}\right]\approx 0,\nonumber \\
\Omega_{10}&=& \pi^{a}_{i}-\epsilon^{0ab}\left[\beta A_{b i}+e_{b i}\right] \approx0.
\label{eq63a}
\end{eqnarray}
in this manner, the matrix whose entries are the Poisson brackets between the constraints, namely $G$,  is given by
\begin{eqnarray}
G(x,y)=
{\tiny
\left(
 \begin{array}{cccccccccccc}
 -2\Omega\epsilon^{0ab}\delta_{ij}&0&0&0&\delta^{j}_{i}\partial_{x}^{a}&-2\epsilon^{0ab}\delta_{ij}&0&0&0&0\\
 0&0&-\delta^{j}_{i}&0&0&0&0&0&0&0\\
 0&\delta^{i}_{j}&0&0&0&0&0&0&0&0\\
 0&0&0&0&\delta^{j}_{i}\nabla_{x}^{2}&0&0&0&0&0\\
 \delta^{i}_{j}\partial_{x}^{b}&0&0&-\delta^{i}_{j}\nabla^{2}_{x}&0&0&0&0&0&0\\
  -2\epsilon^{0ab}\delta_{ij}&0&0&0&0&-2\beta\epsilon^{0ab}\delta_{ij}&0&0&0&\delta^{j}_{i}\partial_{x}^{a}\\
  0&0&0&0&0&0&0&-\delta^{j}_{i}&0&0\\
  0&0&0&0&0&0&\delta^{i}_{j}&0&0&0\\
   0&0&0&0&0&0&0&0&0&\delta^{j}_{i}\nabla^{2}_{x}\\
  0&0&0&0&0&\delta^{i}_{j}\partial_{x}^{b}&0&0&-\delta^{i}_{j}\nabla^{2}_{x}&0\\
  \end{array}
\right)} \nonumber \\
\times \delta^2(x-y).
\label{eq64a}
\end{eqnarray}
Hence,  the inverse of  $G$ becomes

\begin{eqnarray*}
[G(x,y)]^{-1}={\tiny
\left(
 \begin{array}{cccccccccccc}
 \epsilon_{0ab}\delta^{ij}\frac{\beta}{2\theta}&0&0&\epsilon_{0ab}\delta^{i}_{j}\frac{\beta\partial^{b}_{x}}{2\theta\nabla^{2}_{x}}&0&-\epsilon_{0ab}\delta^{ij}\frac{1}{2\theta}&0&0&-\epsilon_{0ab}\delta^{i}_{j}\frac{\partial^{b}_{x}}{2\theta\nabla^{2}_{x}}&0\\
 0&0&\delta^{j}_{i}&0&0&0&0&0&0&0\\
 0&-\delta^{i}_{j}&0&0&0&0&0&0&0&0\\
 \epsilon_{0ba}\delta^{j}_{i}\frac{\beta\partial^{a}_{x}}{2\theta\nabla^{2}_{x}}&0&0&0&-\frac{\delta^{j}_{i}}{\nabla^{2}_{x}}&-\epsilon_{0ba}\delta^{j}_{i}\frac{\partial^{a}_{x}}{2\theta\nabla^{2}_{x}}&0&0&0&0\\
 0&0&0&\frac{\delta^{i}_{j}}{\nabla^{2}_{x}}&0&0&0&0&0&0\\
 -\epsilon_{0ab}\delta^{ij}\frac{1}{2\theta}&0&0&-\epsilon_{0ab}\delta^{i}_{j}\frac{\partial^{b}_{x}}{2\theta\nabla^{2}_{x}}&0&\epsilon_{0ab}\delta^{ij}\frac{\Omega}{2\theta}&0&0&\epsilon_{0ab}\delta^{i}_{j}\frac{\Omega\partial^{b}_{x}}{2\theta\nabla^{2}_{x}}&0\\
 0&0&0&0&0&0&0&\delta^{j}_{i}&0&0\\
 0&0&0&0&0&0&-\delta^{i}_{j}&0&0&0\\
 -\epsilon_{0ba}\delta^{j}_{i}\frac{\partial^{a}_{x}}{2\theta\nabla^{2}_{x}}&0&0&0&0&\epsilon_{0ba}\delta^{j}_{i}\frac{\Omega\partial^{a}_{x}}{2\theta\nabla^{2}_{x}}&0&0&0&-\frac{\delta^{j}_{i}}{\nabla^{2}_{x}}\\
 0&0&0&0&0&0&0&0&\frac{\delta^{i}_{j}}{\nabla^{2}_{x}}&0\
  \end{array}
\right)} \\
\times \delta^2(x-y),
\end{eqnarray*}
here  we have defined  $\theta=\beta\Omega-1$. Finally, we use the inverse matrix $G^{-1}$ and we find the following Dirac's brackets
\begin{eqnarray}
\{e^{i}_{a}(x),p^{b}_{j}(y)\}_{D}&=&\delta{^{i}}_{j}(\delta{^{b}}_{a}-\frac{\partial_{a}\partial^{b}}{\nabla^{2}})\delta(x-y), \nonumber\\
\{e^{i}_{a}(x),e^{j}_{b}(y)\}_{D}&=&0, \nonumber\\
\{p^{a}_{i}(x),p^{b}_{j}(y)\}_{D}&=&0, \nonumber\\
\{A^{i}_{a}(x),\pi^{b}_{j}(y)\}_{D}&=&\delta{^{i}}_{j}(\delta{^{b}}_{a}-\frac{\partial_{a}\partial^{b}}{\nabla^{2}})\delta(x-y), \nonumber\\
\{A^{i}_{a}(x),A^{j}_{b}(y)\}_{D}&=&0, \nonumber \\
\{\pi^{a}_{i}(x),\pi^{b}_{j}(y)\}_{D}&=&0.
\label{eq64f}
\end{eqnarray}
In the following section, we will reproduce these results by mean of  [FJ] formalism.
\section{Faddeev-Jackiw analysis of  [BL] Abelian theory by working with the phase space}
Now, in this section  we shall  study  the action (\ref{eq122})  by means [FJ] formalism, we shall  work   with   the phase space  as symplectic variables \cite{14}.  Hence, the  Lagrangian density can  be written as
\begin{equation}
\mathcal{L}=\epsilon^{0ab}\left[\left(\frac{A^{i}_{b}}{\gamma\sqrt{\mid\Lambda\mid}}+e^{i}_{b}\right)\dot{A}_{ai}+\left(\frac{s\sqrt{\mid\Lambda\mid}}{\gamma}e^{i}_{b}+A^{i}_{b}\right)\dot{e}_{ai}\right]-V^{(0)},
\label{eq81}
\end{equation}
where  the potential symplectic si given by
\begin{equation}
V^{(0)}=-\epsilon^{0ab}\left[\left(\frac{A^{i}_{0}}{\gamma\sqrt{\mid\Lambda\mid}}+e^{i}_{0}\right)F_{abi}+\left(\frac{s\sqrt{\mid\Lambda\mid}}{\gamma}e^{i}_{0}+A^{i}_{0}\right)T_{abi}\right].
\label{eq82}
\end{equation}
By introducing the canonical momenta
\begin{eqnarray}
p^{a}_{i}&=&\epsilon^{0ab}(A_{bi}+\Omega e_{bi}), \nonumber\\
\pi^{a}_{i} &=&\epsilon^{0ab}(e_{bi}+\beta A_{bi}), \nonumber\\
\label{eq83}
\end{eqnarray}
and writing  the fields in the following form
\begin{eqnarray}
\epsilon^{0ab}e_{bi}&=&\frac{s}{s-\gamma^{2}}(\beta p^{a}_{j}-\pi^{a}_{j}), \nonumber\\
\epsilon^{0ab}A_{bi}&=&\frac{s}{s-\gamma^{2}}( \Omega\pi^{a}_{j}-p^{a}_{j}), \nonumber
\label{eq84}
\end{eqnarray}
the first-order symplectic Lagrangian density takes the form
\begin{equation}
\mathcal{L}^{(0)}=\pi^{a}_{i}\dot{A}^{i}_{a}+p^{a}_{i}\dot{e}^{i}_{a}-V^{(0)},
\label{eq86}
\end{equation}
where the potential symplectic $V^{(0)}$ is given by
\begin{equation}
V^{(0)}=-2A^{i}_{0}\partial_{a}\pi^{a}_{i}-2e^{i}_{0}\partial_{a}p^{a}_{i}.
\label{eq87}
\end{equation}
In this manner, we can identify the corresponding symplectic  variables
$\xi{^{(0)a}}(x)=\{e^{i}_{a},p^{a}_{i},e^{i}_{0},A^{i}_{a},\pi^{a}_{i},A^{i}_{0}\} $ and the symplectic 1-form  $a{^{(0)}}{_{a}}(x)=\{p^{a}_{i},0,0,\pi^{a}_{i},0,0\}$,  thus, by using the symplectic variables,    the symplectic matrix takes the form

\begin{eqnarray}
f^{(0)}_{ab}(x,y)=
\left(
 \begin{array}{cccccc}
   0&-\delta{^{a}}_{b}\delta{^{j}}_{i}&0&0&0&0\\
   \delta{^{b}}_{a}\delta{^{i}}_{j}&0&0&0&0&0\\
   0&0&0&0&0&0\\
   0&0&0&0&-\delta{^{a}}_{b}\delta{^{j}}_{i}&0\\
   0&0&0&\delta{^{a}}_{b}\delta{^{i}}_{j}&0&0\\
   0&0&0&0&0&0\\
  \end{array}
\right)\delta^2(x-y),
\label{eq91}
\end{eqnarray}
This matrix is  singular, this means that the theory  has constraints. The zero modes of this matrix are given by  $(v_{a}^{(0)})_{1}^{T}=(0,0,v^{e^{i}_{0}},0,0,0)$ and $(v_{a}^{(0)})_{2}^{T}=(0,0,0,0,0,v^{A^{i}_{0}})$, where $v^{e^{i}_{0}}$ and $v^{A^{i}_{0}}$ are arbitrary functions. Now, by using the zero-modes we can get the following constraints
\begin{eqnarray}
0&=&\int d^{2}x(v^{(0)})^{T}_{a}(x)\frac{\delta}{\delta\xi^{(0)i}(x)}\int d^{2}y V^{(0)}(\xi) \nonumber \\
            &=& \int d^{2}x v^{e^{i}_{0}}(x)[-2\partial_{a}p^{a}_{i}]\nonumber \\
            &\rightarrow&  \Omega^{(0)}_{i}:= -2\partial_{a}p^{a}_{i}=0,
\label{eq92}
\end{eqnarray}

\begin{eqnarray}
0&=&\int d^{2}x(v^{(0)})^{T}_{a}(x)\frac{\delta}{\delta\xi^{(0)i}(x)}\int d^{2}y V^{(0)}(\xi) \nonumber \\
            &=& \int d^{2}x v^{A_{0}}(x)[-2\partial_{a}\pi^{a}_{i}]\nonumber \\
            &\rightarrow&  \Theta^{(0)}_{i}:=-2\partial_{a}\pi^{a}_{i}=0.
\label{eq93}
\end{eqnarray}
On the other hand, we will   research if there are more constraints by calculating the following contraction  \cite{15}
\begin{eqnarray}
   f^{(1)}_{cd}\dot{\xi}^{d}=Z_{c}(\xi),
\label{eq96}
\end{eqnarray}
where
\begin{eqnarray}
f^{(1)}_{cd}=
\left(
 \begin{array}{cccc}
   f^{(0)}_{ab}\\
   \frac{\partial\Omega^{(0)}}{\partial\xi^{a}}\\
  \end{array}
\right)=\left( \begin{array}{cccccc}
   0&-\delta{^{a}}_{b}\delta{^{j}}_{i}&0&0&0&0\\
   \delta{^{b}}_{a}\delta{^{i}}_{j}&0&0&0&0&0\\
   0&0&0&0&0&0\\
   0&0&0&0&-\delta{^{a}}_{b}\delta{^{j}}_{i}&0\\
   0&0&0&\delta{^{b}}_{a}\delta{^{j}}_{i}&0&0\\
   0&0&0&0&0&0\\
   0&2\partial_{b}\delta{^{j}}_{i}&0&0&0&0\\
     0&0&0&0&2\partial_{b}\delta{^{j}}_{i}&0\\
  \end{array}
\right)\delta^2(x-y),
\label{eq97}
\end{eqnarray}
 and
\begin{eqnarray}
Z_{c}(\xi)=
\left(
 \begin{array}{cccc}
   \frac{\partial V^{(0)}(\xi)}{\partial \xi^{a}}\\
   0\\
  \end{array}
\right),
\label{eq98}
\end{eqnarray}
The matrix $(f^{(1)}_{ab})$ given in (\ref{eq97}) is  not obviously a square matrix, but it still has  linearly independent modes given by  $(v^{(1)})_{1}^{T}=(2\partial_{a}v^{\lambda},0,v^{e^{i}_{0}},0,0,0,v^{\lambda},0)$ and $(v^{(1)})_{2}^{T}=(0,0,0,2\partial_{a}v^{\alpha},0,v^{A^{i}_{0}},0,v^{\alpha})$. Multiplication of $(f^{(1)}_{cd})$ by $(v^{(1)})_{c}^{T}$ from the left side gives zero. The contraction of  these modes  reads
\begin{equation}
(v^{(1)})_{c}^{T}Z_{c}\mid_{\Omega^{(0)}=0}=0,
\label{eq100}
\end{equation}
which  is an identity, hence,  there is not new constraints. \\
Furthermore, we use the following  Lagrangian multipliers  $(\lambda^{i},\rho^{i})$ enforcing the constraints  (\ref{eq92}) and  (\ref{eq93})  in order to construct a new symplectic Lagrangian
\begin{equation}
\mathcal{L}^{(1)}=\pi^{a}_{i}\dot{A}^{i}_{a}+p^{a}_{i}\dot{e}^{i}_{a}+(2\partial_{a}p^{a}_{i})\dot{\lambda}^{i}+(2\partial_{a}\pi^{a}_{i})\dot{\rho}^{i}-V^{(1)},
\label{eq103}
\end{equation}
where $V^{(1)}=V^{(0)}\mid_{\partial_{a}\pi^{a}_{i}=0,\partial_{a}p^{a}_{i}=0}=0$,  is  the symplectic potential.  From (\ref{eq103}) we  identify the following symplectic variables
$\xi{^{(1)a}}(x)=\{e^{i}_{a},p^{a}_{i},\lambda^{i},A^{i}_{a},\pi^{a}_{i},\rho^{i}\}, $ and the 1-forms $a{^{(1)}}{_{a}}(x)=\{p^{a}_{i},0,2\partial_{a}p^{a}_{i},\pi^{a}_{i},0,2\partial_{a}\pi^{a}_{i}\} $, thus, the corresponding symplectic matrix is given by
\begin{eqnarray}
f^{(1)}_{ab}(x,y)=
\left(
 \begin{array}{cccccc}
   0&-\delta{^{a}}_{b}\delta{^{j}}_{i}&0&0&0&0\\
   \delta{^{b}}_{a}\delta{^{i}}_{j}&0&-2\delta{^{i}}_{j}\partial^{}_{a}&0&0&0\\
   0&-2\delta{^{j}}_{i}\partial^{}_{b}&0&0&0&0\\
   0&0&0&0&-\delta{^{a}}_{b}\delta{^{j}}_{i}&0\\
   0&0&0&\delta{^{b}}_{a}\delta{^{i}}_{j}&0&-2\delta{^{i}}_{j}\partial^{}_{a}\\
   0&0&0&0&-2\delta{^{j}}_{i}\partial^{}_{b}&0\\
  \end{array}
\right)\delta^2(x-y),
\label{eq105}
\end{eqnarray}
the  matrix   is still singular, but we have proved, however, that  there are not   new constraints. Therefore this system has a gauge symmetry. In oder to obtain a symplectic tensor, we will fix the gauge,  let us fixing the Coulomb gauge $\partial^{a}e^{i}_{a}=0$, $\partial^{a}A^{i}_{a}=0$  and we will introduce this information by constructing a new symplectic  Lagrangian adding new Lagrange multiples,  namely $\phi_{i}$ and $\theta_{i}$, enforcing the gauge fixing, we obtain
\begin{equation}
\mathcal{L}^{(2)}=\pi^{a}_{i}\dot{A}^{i}_{a}+p^{a}_{i}\dot{e}^{i}_{a}+(2\partial_{a}p^{a}_{i})\dot{\lambda}^{i}+(2\partial_{a}\pi^{a}_{i})\dot{\rho}^{i}+(\partial^{a}e^{i}_{a})\dot{\phi_{i}}+(\partial^{a}A^{i}_{a})\dot{\theta_{i}},
\label{eq106}
\end{equation}
now the symplectic variables are given by  $\xi{^{(2)}}(x)=\{e^{i}_{a},p^{a}_{i},\lambda^{i},\phi_{i},A^{i}_{a},\pi^{a}_{i},\rho^{i},\theta_{i}\}$ and the 1-forms  $a{^{(2)}}{_{}}(x)=\{p^{a}_{i},0,2\partial_{a}p^{a}_{i},\partial^{a}e^{i}_{a},\pi^{a}_{i}, 0,2\partial_{a}\pi^{a}_{i},\partial^{a}A^{i}_{a}\}$. In this manner, the symplectic matrix takes the form
\begin{eqnarray*}
f^{(2)}_{ab}(x,y)=
\left(
 \begin{array}{cccccccc}
   0&-\delta{^{a}}_{b}\delta{^{j}}_{i}&0&-\delta{^{j}}_{i}\partial^{}_{a}&0&0&0&0\\
   \delta{^{b}}_{a}\delta{^{i}}_{j}&0&-2\delta{^{i}}_{j}\partial^{}_{a}&0&0&0&0&0\\
   0&-2\delta{^{j}}_{i}\partial^{}_{b}&0&0&0&0&0&0\\
   -\delta{^{i}}_{j}\partial^{}_{b}&0&0&0&0&0&0&0\\
   0&0&0&0&0&-\delta{^{a}}_{b}\delta{^{j}}_{i}&0&-\delta{^{j}}_{i}\partial^{}_{a}\\
   0&0&0&0&\delta{^{b}}_{a}\delta{^{i}}_{j}&0&-2\delta{^{i}}_{j}\partial^{}_{a}&0\\
   0&0&0&0&0&-2\delta{^{j}}_{i}\partial^{}_{b}&0&0\\
   0&0&0&0&-\delta{^{i}}_{j}\partial^{}_{b}&0&0&0\\
  \end{array}
\right)
\end{eqnarray*}

\begin{eqnarray}
\times\delta^2(x-y),
\label{eq108}
\end{eqnarray}
where we can observe that $f^{(2)}_{ab}(x,y)$  is an symplectic tensor and therefore   is invertible. The inverse matrix of $f^{(2)}_{ab}(x,y )$ is given by
\begin{eqnarray*}
[f^{(2)}_{ab}(x,y)]^{-1}= \nonumber\\
\end{eqnarray*}

\begin{eqnarray*}
\left(
 \begin{array}{cccccccc}
   0&\delta{^{i}}_{j}(\delta{^{b}}_{a}-\frac{\partial_{a}\partial^{b}}{\nabla^{2}})&0&-\delta{^{i}}_{j}\frac{\partial_{a}}{\nabla^{2}}&0&0&0&0\\
   -\delta{^{j}}_{i}(\delta{^{a}}_{b}-\frac{\partial_{b}\partial^{a}}{\nabla^{2}})&0&-\frac{1}{2}\delta{^{j}}_{i}\frac{\partial^{a}}{\nabla^{2}}&0&0&0&0&0\\
   0&-\frac{1}{2}\delta{^{i}}_{j}\frac{\partial^{b}}{\nabla^{2}}&0&-\delta{^{i}}_{j}\frac{1}{2}\frac{1}{\nabla^{2}}&0&0&0&0\\
   -\delta{^{j}}_{i}\frac{\partial_{b}}{\nabla^{2}}&0&\frac{1}{2}\delta{^{j}}_{i}\frac{1}{\nabla^{2}}&0&0&0&0&0\\
    0&0&0&0&0&\delta{^{i}}_{j}(\delta{^{b}}_{a}-\frac{\partial_{a}\partial^{b}}{\nabla^{2}})&0&-\delta{^{i}}_{j}\frac{\partial_{a}}{\nabla^{2}}\\
   0&0&0&0&-\delta{^{j}}_{i}(\delta{^{a}}_{b}-\frac{\partial_{b}\partial^{a}}{\nabla^{2}})&0&-\frac{1}{2}\delta{^{j}}_{i}\frac{\partial^{a}}{\nabla^{2}}&0\\
   0&0&0&0&0&-\frac{1}{2}\delta{^{i}}_{j}\frac{\partial^{b}}{\nabla^{2}}&0&-\frac{1}{2}\delta{^{i}}_{j}\frac{1}{\nabla^{2}}\\
   0&0&0&0&-\delta{^{j}}_{i}\frac{\partial_{b}}{\nabla^{2}}&0&\frac{1}{2}\delta{^{j}}_{i}\frac{1}{\nabla^{2}}&0\\
  \end{array}
\right)
\end{eqnarray*}
\begin{eqnarray}
\times\delta^2(x-y)
\label{eq109}
\end{eqnarray}
where it is possible identify the [FJ] generalized brackets as
\begin{eqnarray}
\{\xi_{a}^{(2)}(x),\xi_{b}^{(2)}(y)\}_{FD}=[f^{(2)}_{ab}(x,y)]^{-1},
\end{eqnarray}
thus, we find the following brackets
\begin{eqnarray}
\{e^{i}_{a}(x),p^{b}_{j}(y)\}_{FD}=[f^{(2)}_{12}(x,y)]^{-1}&=&\delta{^{i}}_{j}(\delta{^{b}}_{a}-\frac{\partial_{a}\partial^{b}}{\nabla^{2}})\delta^2(x-y), \nonumber \\
\{A^{i}_{a}(x),\pi^{b}_{j}(y)\}_{FD}=[f^{(2)}_{56}(x,y)]^{-1}&=&\delta{^{i}}_{j}(\delta{^{b}}_{a}-\frac{\partial_{a}\partial^{b}}{\nabla^{2}})\delta^2(x-y), \nonumber \\
\{e^{i}_{a}(x),e^{j}_{b}(y)\}_{FD}=[f^{(2)}_{11}(x,y)]^{-1}&=&0,\nonumber \\
\{A^{i}_{a}(x),A^{j}_{b}(y)\}_{FD}=[f^{(2)}_{55}(x,y)]^{-1}&=&0, \nonumber \\
\{p^{a}_{i}(x),p^{b}_{j}(y)\}_{FD}=[f^{(2)}_{22}(x,y)]^{-1}&=&0, \nonumber \\
\{\pi^{a}_{i}(x),\pi^{b}_{j}(y)\}_{FD}=[f^{(2)}_{66}(x,y)]^{-1}&=&0, \nonumber \\
\{p^{a}_{i}(x),\lambda^{j}(y)\}_{FD}=[f^{(2)}_{23}(x,y)]^{-1}&=&-\frac{1}{2}\delta^{j}_{i}\frac{\partial^{a}}{\nabla^{2}}\delta^2(x-y), \nonumber \\
\{\pi^{a}_{i}(x),\rho^{j}(y)\}_{FD}=[f^{(2)}_{67}(x,y)]^{-1}&=&-\frac{1}{2}\delta^{j}_{i}\frac{\partial^{a}}{\nabla^{2}}\delta^2(x-y), \nonumber \\
\{\lambda^{i}(x),\phi_{j}(y)\}_{FD}=[f^{(2)}_{34}(x,y)]^{-1}&=&-\frac{1}{2}\delta^{i}_{j}\frac{1}{\nabla^{2}}\delta^2(x-y), \nonumber \\
\{\rho^{i}(x),\theta_{j}(y)\}_{FD}=[f^{(2)}_{78}(x,y)]^{-1}&=&-\frac{1}{2}\delta^{i}_{j}\frac{1}{\nabla^{2}}\delta^2(x-y), \nonumber \\
\{e^{i}_{a}(x),\phi_{j}(y)\}_{FD}=[f^{(2)}_{14}(x,y)]^{-1}&=&-\delta^{i}_{j}\frac{\partial_{a}}{\nabla^{2}}\delta^2(x-y), \nonumber \\
\{A^{i}_{a}(x),\theta_{j}(y)\}_{FD}=[f^{(2)}_{58}(x,y)]^{-1}&=&-\delta^{i}_{j}\frac{\partial_{a}}{\nabla^{2}}\delta^2(x-y).
\label{eq121}
\end{eqnarray}
We can observe that the  Dirac brackets given in (\ref{eq64f}) and the [FJ] generalized brackets  given in (\ref{eq121})  coincide to each other.\\



\end{document}